\documentclass{article}

\usepackage[final]{styles/neurips_2024}

\usepackage[utf8]{inputenc} 
\usepackage[T1]{fontenc}    
\usepackage{hyperref}       
\usepackage{url}            
\usepackage{booktabs}       
\usepackage{amsfonts}       
\usepackage{nicefrac}       
\usepackage{microtype}      
\usepackage{xcolor}         

\usepackage{hyperref}
\usepackage{url}
\usepackage{graphicx}
\usepackage{doi}        
\usepackage{amsmath}
\usepackage{pifont}
\usepackage{xcolor}
\usepackage{bm}

\usepackage{algorithm}
\usepackage{algpseudocode}

\newcommand{\model}{\textit{Non-transitive Active Markov Game Model}}

\newcommand{\feint}{Feint\ }

\title{\feint Behaviors and Strategies:\\Formalization, Implementation and Evaluation}

\author{%
  Junyu Liu \\
  Brown University\\
  \texttt{ljunyu381@gmail.com} \\
  \And
  Xiangjun Peng \\
  The Chinese University of Hong Kong \\
  \texttt{xjpeng@cse.cuhk.edu.hk} \\
}

\begin{document}

\maketitle

\begin{abstract}
  \feint behaviors refer to a set of deceptive behaviors in a nuanced manner, which enable players to obtain temporal and spatial advantages over opponents in competitive games. Such behaviors are crucial tactics in most competitive multi-player games (e.g., boxing, fencing, basketball, motor racing, etc.). However, existing literature does not provide a comprehensive (and/or concrete) formalization for \feint behaviors, and their implications on game strategies. In this work, we introduce the first comprehensive formalization of \feint behaviors at both action-level and strategy-level, and provide concrete implementation and quantitative evaluation of them in multi-player games. The key idea of our work is to (1) allow automatic generation of \feint behaviors via Palindrome-directed templates, combine them into meaningful behavior sequences via a Dual-Behavior Model; (2) concertize the implications from our formalization of \feint on game strategies, in terms of temporal, spatial and their collective impacts respectively; and (3) provide a unified implementation scheme of \feint behaviors in existing MARL frameworks. The experimental results show that our design of \feint behaviors can (1) greatly improve the game reward gains; (2) significantly improve the diversity of Multi-Player Games; and (3) only incur negligible overheads in terms of time consumption.
\end{abstract}

\vspace{-8pt}

\section{Introduction}

\vspace{-8pt}

In most real-world Multi-Player Games (e.g., boxing, basketball, motor racing, and etc.), players exhibit complex behaviors, which imply hard-to-be-quantified interactions. Simulating these games requires to model the players' behaviors into action spaces as the elements (i.e., which we denote such a decision as ``action-level"), and explore strategies based on these elements~\cite{SIGGRAPH10/Animation-Two-Player-Game, SIGGRAPH21/Two-Player-Game}. As one of the most common behaviors in real-world games, \feint behaviors represent a class of tactic behaviors, which are used to mislead opponents to obtain temporally-strategic advantages\footnote{Note that the resulting advantages may be exhibited temporally and/or spatially.}. Such behaviors generally exhibit nuanced differences with normal behaviors, in terms of action movements\footnote{Examples include fake punch in boxing, early-braking in motor racing, and etc.}; but these behaviors can help the agent to obtain strategic advantages to a considerable extent\footnote{And, clearly, \feint behaviors can also increase the diversity of those games, according to \cite{NIPS21/BD-RD, AAMAS20/Is-PG-Gradient}.}. However, \underline{no} existing literature provides a comprehensive formalization (and/or concrete modeling of) \feint behaviors, at either action level or strategy level, and we justify this claim below.

\vspace{-8pt}

\begin{itemize}
    \item To the best of our knowledge, \cite{SIGGRAPH10/Animation-Two-Player-Game} is the first work to mention \feint behaviors, as a proof-of-concept to construct animations for nuanced behaviors in two-player boxing games, based on the unpredictability introduced in stochastic and simultaneous decision making. 
    \item A more recent work \cite{SIGGRAPH21/Two-Player-Game} learn control strategies to animate nuanced agent behaviors like \feint in gaming environments, from motion clips using Deep Reinforcement Learning.
\end{itemize}

\textbf{\underline{Summary.}} Prior attempts do not provide detailed formalization to address the action-level characteristic of \feint and do not give \feint behavior generation guidelines. As for the strategy-level formalization, existing learning-based works either neglect \feint behaviors or assume that \feint are just glitches from other behaviors, by inducing similar impacts implicitly.\footnote{As shown in this work, existing learning-based approaches can not effectively model \feint behaviors in strategy-level, since \feint behaviors require intricate planning which is an active process.}

\underline{\textbf{Overview.}}
Our work provides the first comprehensive and concrete formalization of \feint behaviors in action-level and strategy-level. We first present an automatic approach to generate \feint behaviors using \textbf{Palindrome-directed Templates} based on our observation on \feint characteristics, and provide \textbf{Dual-Behavior Model} to examine the design for the combination of \feint behaviors and normal actions. Based on the action-level formalization, we then model the \feint behavior impacts on strategy-level in terms of the temporal, spatial, and their collective impacts under a learnable scheme. Then, we provide a concrete and unified implementation to incorporate the action-level and strategy-level formalization in common Multi-Player Reinforcement Learning (MARL) frameworks; so that we can showcase the effectiveness of our formalization\footnote{To deliver a unified definition of \feint behavior in both continuous and discrete action space, we highlight the difference in appendix~\ref{appendix:action-behavior-dif}}. 

\underline{\textbf{Results.}} To properly examine the effectiveness of our formalization, we extensively construct a complex and physics-based boxing game as an abstraction of some animation-related works~\cite{SIGGRAPH10/Animation-Two-Player-Game, SIGGRAPH21/Two-Player-Game}. We use a two-player and a six-player scenario with 4 commonly used MARL models (MADDPG~\cite{NIPS17/MADDPG}, MASAC~\cite{ICML18/SAC, ICML19/MAAC}, MATD3~\cite{NIPS19/MATD3}, and MAD3PG~\cite{ICLR18/D4PG,MDPI21/MAD3PG}) to extensively evaluate our formalization. We also evaluate our formalization of \feint in a strategic real-game, Alpha Star, to examine the diversity gain introduced by our formalization. The results show that our formalization of \feint can significantly increase the gaming rewards in all scenarios with all 4 MARL models. We also extensively examine our approaches: for the Diversity Gain, our method can increase the exploitation of the search space by 1.98X, measured by the Exploitability metrics; and our implementation scheme only incur less than 5\% overheads in terms of per game episode time consumption. We conclude that our formalization of \feint behaviors is effective and practical, significantly increasing players' game rewards and making Multi-Player Games more interesting.

\vspace{-12pt}

\section{Background}

\vspace{-8pt}

\subsection{\feint Behaviors in the Real-World and Simulated Games}

\vspace{-8pt}

\feint behaviors are common for human players, as a set of active actions to obtain strategic advantages in real-world games. Examples can include sports games such as boxing, basketball, and motor racing \cite{Feint-example-1, Feint-example-2, Feint-example-3}, and electronic games such as King of Fighters and Starcraft \cite{Feint-example-KingOfFighters, Feint-example-Starcraft}. \feint behaviors are not simple deceptive behaviors as their goal is to not to gain rewards for themselves but to create temporal and spatial advantages for some short-term follow-up actions. In addition, \feint behaviors have nuanced action formalizations. Though \feint is undoubtedly important in many real-world games, there still lacks a comprehensive formalization of \feint in Multi-Player Game simulations using Non-Player Characters (NPCs). There are only a limited amount of works to tackle this issue. \cite{SIGGRAPH10/Animation-Two-Player-Game} is an early example of incorporating \feint as a proof-of-concept, which focuses on constructing animations for nuanced game strategies for more unpredictability from NPCs. More recently, \cite{SIGGRAPH21/Control-Strategy-Two-Player-Game} learn multi-level control strategies of agents via deep reinforcement learning from a set of motion clips including nuanced behaviors like \feint (i.e. in animating combat scenes). However, these prior works (1) lack concrete formalizations of \feint behavior characteristics, which cannot fully unveil the variety of \feint behaviors in the action level; and (2) lack comprehensive explorations of \feint behaviors implications on game strategies, which neglects the potential impacts of fusing effective \feint behaviors into strategies; and (3) solely focus on Two-Player Games, which can not be effectively generalized to multi-player scenarios.

\vspace{-8pt}

\subsection{MARL Models at Strategy-Level in Multi-Player Game Simulations}

\vspace{-8pt}

Multi-Agent Reinforcement Learning (MARL) aims to learn optimal policies for agents in a multi-agent environment, which consists of various agent-agent and agent-environment interactions\footnote{Note that these efforts can establish \feint upon prior arts (as covered in appendix~\ref{appendix:feint-action}), and we have justified the novelty of our approach in appendix~\ref{appendix:feint-action}.}. Many single-agent Reinforcement Learning methods (e.g. DDPG~\cite{ICLR2016/DDPG}, SAC~\cite{ICML18/SAC}, PPO~\cite{Journal17/PPO} and TD3~\cite{ICML18/TD3}, D4PG~\cite{ICLR18/D4PG}) can not be directly used in multi-agent scenarios, since the rapidly-changing multi-agent environment can cause highly unstable learning results (evidenced by~\cite{NIPS17/MADDPG}). Thus, recent efforts on MARL model designs aim to address such an issue. \cite{AAAI18/COMA} proposes Counterfactual Multi-Agent (COMA) policy gradients, which uses centralised critic to estimate the Q-function and decentralised actors to optimize agents’ policies. \cite{NIPS17/MADDPG} proposes Multi-Agent Deep Deterministic Policy Gradient (MADDPG), which decreases the variance in policy gradient and instability of Q-function of DDPG in multi-agent scenarios. \cite{ICML19/MAAC} proposes Multi-Agent Actor-attention Critic (MAAC), which applies attention entropy mechanism to enable effective and scalable policy learning. These models can have varied impacts within a diverse set of scenarios. \cite{MDPI21/MAD3PG} introduces Multi-agent Distributed Deep Deterministic Policy Gradient (MAD3PG), which extends the D4PG to multi-agent scenarios with distributed critics to enable distributed tracking. \cite{NIPS19/MATD3} proposes Multi-Agent Twin Delayed Deep Deterministic Policy Gradient (MATD3), which integrates twin delayed Q-learning and addressing the overestimation bias in Q-values in a multi-agent setting. Though different MARL models have different design details, they all share the same high-level learning structure. Thus, our goal is to provide a unified scheme to fuse our formalization of \feint behaviors into game simulations that can be learned using common MARL models, enabling effective \feint behaviors impacts regardless of specific design choices of MARL models. 

\vspace{-12pt}

\section{Formalizing \feint Behavior}
\label{sec:formalization-action-level}

\vspace{-8pt}

We introduce our formalization of \feint behaviors in action level regarding (1) how to automatically generate \feint behavior with templates from common attack behaviors; and (2) how can the generated \feint behaviors be synergistically combined with follow-up high-reward actions. We first introduce our methodology to automatically generate \feint behaviors, by exploiting our newly-revealed insight called \textbf{Palindrome-directed Generation of \feint Templates}. Next, we illustrate key design choices on how to combine the generated \feint behaviors with follow-up actions in a \textbf{Double-Behavior Model}, which forms the foundation for the designs of Feint-accounted strategy designs in Section~\ref{sec:formalization-strategy-level}\footnote{We choose boxing game as an example to concretely explain our insights for \feint behaviors in this section but our formalization is a unified abstraction of common games and can be easily adapted to other games including basketball, fencing, motor racing, etc.}.

\vspace{-8pt}

\subsection{\feint Behavior Characteristics and Templates}
\label{subsec:action-Feint-templates}
\vspace{-8pt}

\begin{figure*}[!h]
  \centering
\includegraphics[width=\linewidth]{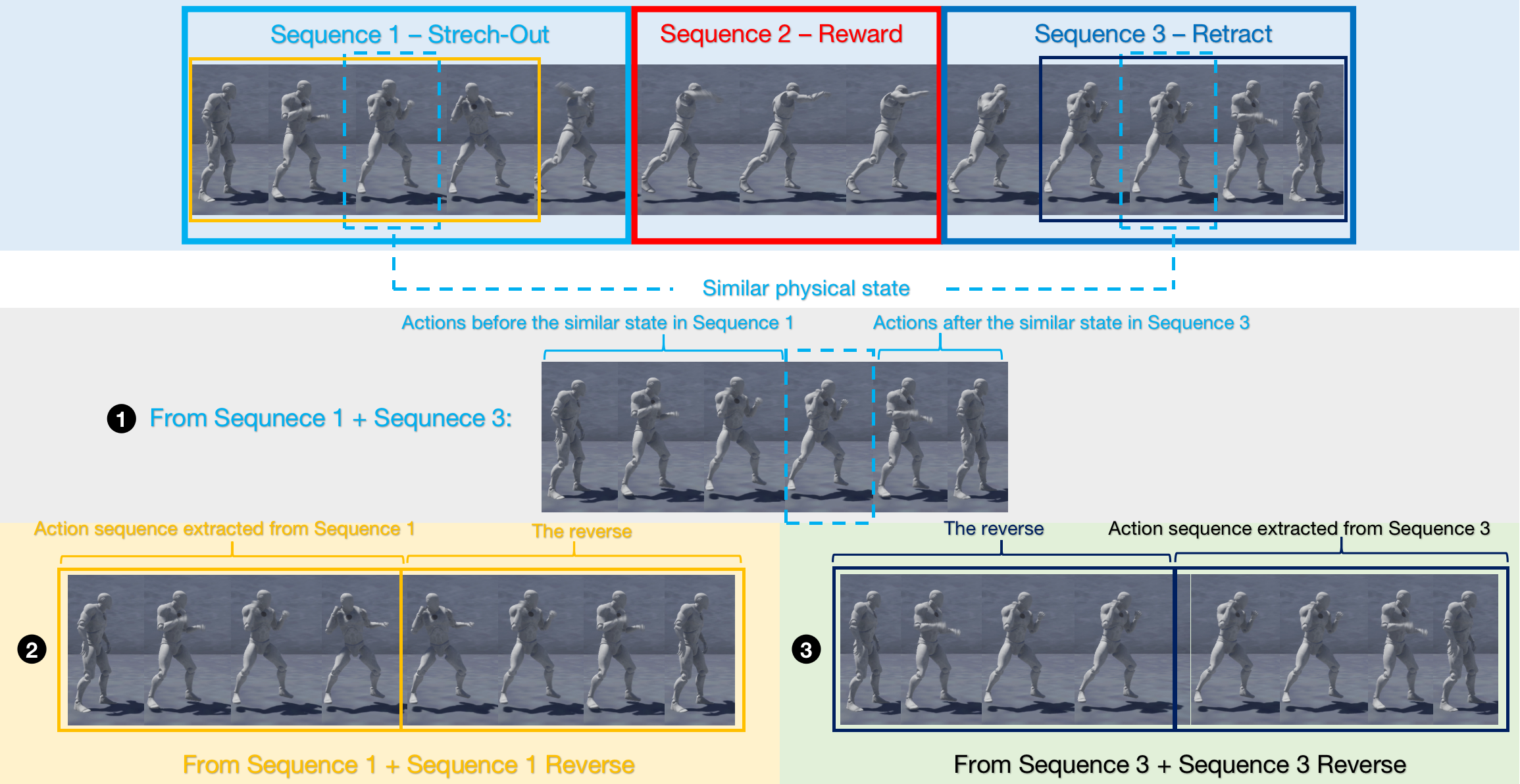}
  \caption{An example of \textbf{Palindrome-directed Generation Templates of \feint behaviors}. The first row shows an action sequence of a cross-punch behavior. Three examples of templates are shown as \ding{202}, \ding{203}, and \ding{204} to demonstrate physically realistic generation of \feint behaviors.}
  \label{fig:three-stage}
\end{figure*}

Since \feint behaviors aim to provide deceptive attacks, they are naturally expected to be derived from a subset of existing attack behaviors. Based on our exploration, we derive two key findings from an extensive amount of attack behaviors. First, most attack behaviors can be decomposed into three action sequences, which are Stretch-out Sequence (Sequence 1), Reward Sequence (Sequence 2), and Retract Sequence (Sequence 3) (an example shown in the first row in Figure~\ref{fig:three-stage}). We elaborate on each action sequence in detail. 

Sequence 1 leads the agent's movements to the Reward Sequence (in boxing, approaching the opponents before actually hitting them); Sequence 2 contains actions that gain game rewards (in boxing, physical contact with the opponents); and Sequence 3 retracts an agent's movements to a relative rest position (in boxing, retracting back to a preparation position for next behaviors). Second, body movements in Sequence 1 and Sequence 3 usually have semi-symmetric yet reverse-order action patterns in the timeline. A behavior usually starts and ends in a similar physical state due to physical restrictions (e.g., bones and muscles stretching restrictions for a humanoid).

The above three-stage decomposition of attack behaviors has motivated a series of constraints, to deliver proper design of \feint generators. To satisfy the above two requirements, we propose a \feint behavior template generator called \textbf{Palindrome-directed Generation of \feint Templates}, by extracting subsets of semi-symmetrical actions from an attack behavior and synthesizing them as a \feint behavior. The general method to generate these templates are (1) by extracting subsets of unit actions from an attack behavior, a \feint behavior can be considered as a semi-finished real attack behavior. This ensures the high similarity of a generated \feint behavior with an attack behavior, thus opponents can be deceived; and (2) by synthesizing semi-symmetric action sections, the overall movements can be connected smoothly and the naturalness of humanoid actions can be guaranteed. Within our proposed template generator \textbf{Palindrome-directed Generation of \feint Templates}, there are two key adjustable parameters in practice: (1) sequence composition positions for \feint templates; and (2) sequence length for \feint templates. We describe our rationale in appendix~\ref{appendix:feint-actions-3-stage}. Figure~\ref{fig:three-stage} demonstrates 3 templates for generating \feint behavior templates in boxing games.
\vspace{-8pt}

\subsection{\feint Behavior in Consecutive Game Steps}
\label{subsec:action-Dual-Behavior-Model}
\vspace{-8pt}

Standalone \feint behaviors are meaningless in competitive games since the \feint behaviors themselves do not gain rewards. Only by effectively combining \feint behaviors with intended follow-up actions can showcase their effectiveness. Thus, we define an effective \feint cycle as a \textbf{Dual-Behavior Model}, which jointly considers a \feint behavior and its intended follow-up behavior (can be a single action or an action sequence). Our formalization for standalone \feint behaviors (Section~\ref{subsec:action-Feint-templates}) already provides a large number of possible \feint behaviors. However, not all these morphologically reasonable \feint actions can be directly combined with all high-reward follow-up actions in combating scenarios. Therefore, certain constraints are demanded to construct effective combinations of \feint behaviors and follow-up actions. Hereby, we introduce two major considerations and then propose relevant restrictions, to enable naturalistic and suitable combinations of \feint behaviors and follow-up actions.

\begin{figure}[h]
  \centering
\includegraphics[width=\linewidth]{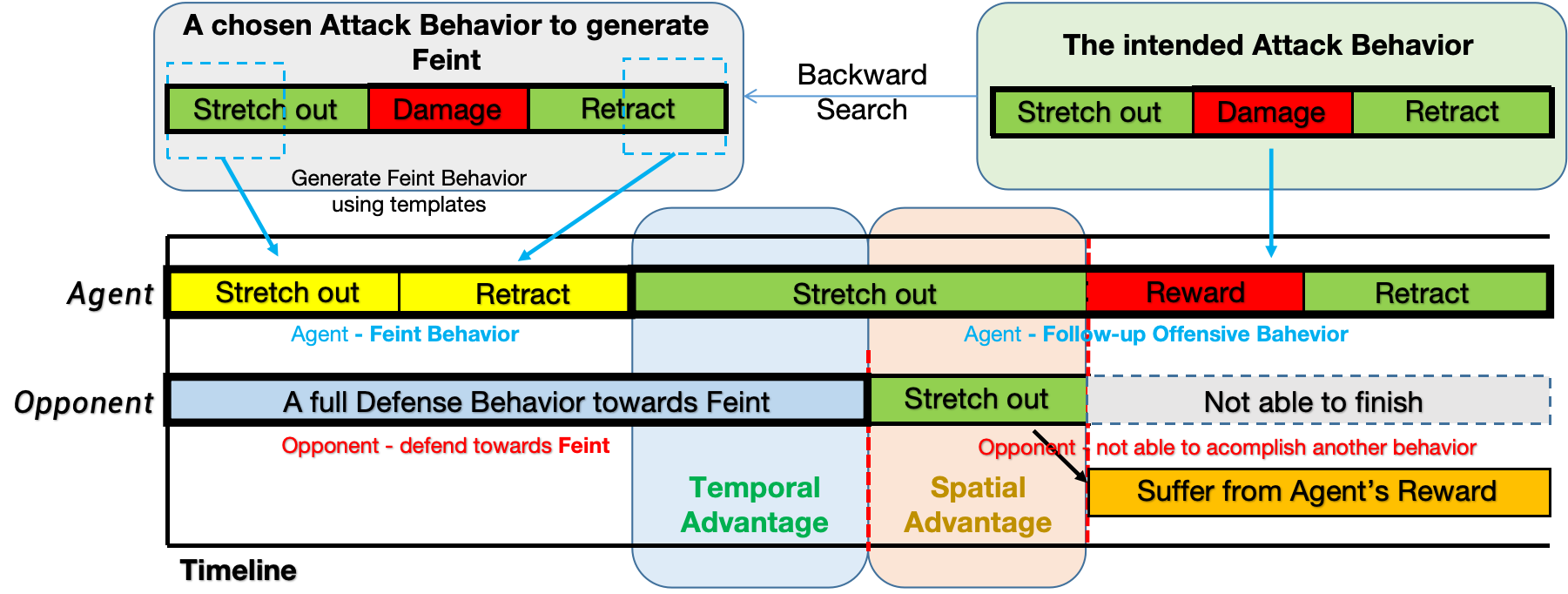}
  \caption{Dual-action Model - high-level abstraction and demonstration of internal stage transitions}
  \label{fig:dual-behavior-model-abstract}
\end{figure}

(1) \textbf{Physical Constraints:} Physical constraints need to be accounted for when synthesizing \feint behaviors and follow-up actions. 
The ending physical state for a \feint behavior must be a state that is physically possible for an agent to perform the follow-up high-reward actions. For example, if an agent in boxing games finishes \feint behavior with the left foot forward, but the following attack behavior starts with the right foot forwarded, the synthesis of these two behavior is inappropriate since this combination is physically unrealistic.

To ensure that the combinations of \feint behaviors and follow-up actions obey the physical constraints, we use a Reverse Search Principle which decides the intended follow-up actions (behavior) first and then use the starting physical state of this behavior to search and compose proper \feint behaviors (a more detailed description combined with strategy is described in Section~\ref{sec:implementation}). By first selecting an intended follow-up high-reward behavior, the end physical state of the \feint behavior is constrained to be close to the starting physical state of the follow-up behavior. Thus the composition of possible \feint behaviors using the Palindrome-directed templates should aim to start and end at a physical state that is close to the follow-up behavior.

(2) \textbf{Effectiveness:} The effectiveness of the incorporation of \feint behaviors is evaluated by whether the following attack actions can successfully gain rewards from the opponent. A successful \feint behavior would usually enable an agent to gain temporal and spatial advantages when performing the follow-up behaviors. Thus, the two design parameters introduced in Section~\ref{subsec:action-Dual-Behavior-Model} play crucial roles in combining \feint with follow-up behaviors. The abstraction of an ideal Dual-Behavior model that can enable an agent with temporal and spatial advantage is illustrated in Figure~\ref{fig:dual-behavior-model-abstract} and a corresponding example is provided in Figure~\ref{fig:dual-behavior-model-example}. An effective \feint behavior creates temporal advantages that make the opponents to defend in a wrong direction and enable temporal advantages to allow the follow-up high-reward behavior to successfully gain rewards on the opponents. 

To ensure the consistency and correctness of the understanding, we provide a detailed demonstration for successful and unsuccessful \feint cases in Appendix~\ref{appendix:feint-demo}.

\vspace{-12pt}

\section{Formalizing \feint Behaviors in Strategies}
\label{sec:formalization-strategy-level}

\vspace{-8pt}

To effectively fuse the \feint beahviors with Dual-Behavior Model into game interactions, we provide the strategy-level formalization of \feint behaviors. We use Multi-Agent Reinforcement Learning (MARL) schemes to discuss our formalization of \feint behaviors in the strategy level, as MARL provides flexibility in exposing multiple adjustable parameters in learnable policy models. As discussed in generating \feint behaviors (Section~\ref{subsec:action-Feint-templates}) and composing them in the Dual-Behavior Models (Section~\ref{subsec:action-Dual-Behavior-Model}), the key considerations for effective \feint cycle is to enable temporal and spatial advantages for an agent. Thus, our strategy-level formalization centers on how to address the temporal, spatial, and their collective impacts of \feint behaviors with Dual-Behavior Models. A more concrete introduction for fusing of \feint into the MARL frameworks is presented in Section~\ref{sec:implementation}.

\vspace{-8pt}

\subsection{The Basic Formalization: Derivation and Limitations}
\label{subsec:strategy-formalization-limitations}

\vspace{-8pt}

\noindent
Under commonly used MARL schemes~\cite{ICML21/BD, NIPS21/BD-RD}, we define a $K$-agent \model\  as a tuple $\left \langle K, \mathcal{S}, \mathcal{A}, P, R, \Pi, \Theta \right \rangle$: $\mathcal{S}$ is the state space; $\mathcal{A}=\{\mathcal{A}_i\}_{i=1}^{K}$ is the set of action space for each agent, where there are no dominant actions; $P$ performs state transitions of current state by agents' actions: $P: \mathcal{S} \times \mathcal{A}_1 \times \mathcal{A}_2 \times ... \times \mathcal{A}_K \rightarrow \mathcal{P}(\mathcal{S})$, where $\mathcal{P}(\mathcal{S})$ denotes the set of probability distribution over state space $S$; $R=\{R_i\}_{i=1}^{K}$ is the set of reward functions for each agent; $\Pi=\{\pi_i\}_{i=1}^{K}$ is the set of policy models for each agent; and $\Theta=\{\Theta_i\}_{i=1}^{K}$ is the set of policy parameters for each agent's policy model. For simplicity, we use $i$ to denote an agent from our description perspective and use $-i$ to refer to all other agents.

We first summarize two major limitations of existing works to justify that they cannot deliver a sufficient formalization of \feint behaviors in Multi-Player Games. Since there is no prior formalization, we discuss relevant works and derive the key features to discuss them in detail. 

\ding{202} The basic formalization on temporal impacts is insufficient for Multi-Player Games. Multi-Player Games require agents to account for complex future planning for decision-making, which is critical for deceptive behaviors like \feint~\cite{NIPS14/DQN, NIPS20/Discounted-not-Optimal, AAMAS20/Is-PG-Gradient}. Several works simplify the temporal impacts of deceptive game strategies in different gaming scenarios. \cite{NIPS14/DQN} uses a discount factor $\gamma$ to calculate the reward for an agent $i$ following actions as $\sum_{t=0}^\infty \gamma^t R^i(s_t, a_t^i, a_t^{-i})$ for agent $i$. However, such a method suffers from the "short-sight" issue \cite{NIPS20/Discounted-not-Optimal}, since the weights for future actions’ rewards shrink exponentially with time, which are not suitable for all gaming situations (discussed in~\cite{AAMAS20/Is-PG-Gradient}). More recently, \cite{ICLR22/Active-Markov-Game} applies a long-term average reward for an agent $i$, to equalize the rewards of all future actions as $\frac{1}{T} \sum_{t=0}^TR^i(s_t, a_t^i, a_t^{-i})$. However, such a method is restricted by the "far-sight" issue, since there are no differentiation between near-future and far-future planning. The mismatch between abstraction granularity heavily saddles with the design of \feint, because they use relatively static representations (e.g. static $\gamma$ and $T$). Therefore, they cannot be aware of any potential changes of strategies in different phases of a game. Hence, the temporal dimension is simplified hereby.

\ding{203} The basic formalization of spatial impacts are generally in simplified 2-player scenarios only, which cannot be effectively generalized to Multi-Player Game scenarios. Prior works, which attempt to fuse \feint into complete game scenarios, only consider two-player scenarios~\cite{SIGGRAPH21/Control-Strategy-Two-Player-Game, Journal22/Entropy-Game}. However, in Multi-Player (more then two player) Games, gaming strategies (especially deceptive strategies) yield spatial impacts on other agents. Such impacts have been overlooked by all prior works. This is because an agent, who launches the \feint behaviors, can impact not only the target agent but also other agents in the scenario. Therefore, the influences of such an action needs to account for spatial impacts~\cite{NIPS21/BD-RD}. Moreover, with a new dimension accounted, the interactions between them also raise a potential issue for their mutually collective impacts. 
\vspace{-8pt}

\subsection{Our Formalization in a Generalized Game Model}
\vspace{-8pt}

Therefore, to deliver an effective formalization of \feint in Multi-Player Games, it is essential to consider the temporal, spatial and their collective impacts comprehensively. We first discuss the Temporal Dimension, then we elaborate our considerations on Spatial Dimension, and finally we summarize the design for the collective impacts from both temporal and spatial dimensions.

\vspace{-8pt}

\subsubsection{Temporal Dimension: Influence Time}
\label{subsubsec:strategy-temporal}
\vspace{-8pt}

To formalize the temporal impacts of \feint behaviors based on our Palindrome-directed Templates and the Dual-Behavior Model, we use a \textit{Dynamic Short-Long-Term} manner to emulate them, which differ from the prior works' formalization (Section~\ref{subsec:strategy-formalization-limitations}). The short-term period refers to a complete Dual-Behavior Model (Section~\ref{subsec:action-Dual-Behavior-Model}), including a \feint behavior followed by an intended high-reward behavior led by the Feint. The long-term period is the time steps after this \feint cycle. The rationale behind such a design choice is that: the purpose of \feint is to obtain strategic advantages against the opponent in the temporal dimension, aiming to benefit the follow-up high-reward behavior. Hence, the \textit{Dynamic Short-Long-Term} temporal impacts of \feint shall be (1) the actions that follow \feint beahviors (e.g. actual attacks) in a short-term period of time should have a strong correlation to \feint; (2) the actions in the long-term periods explicitly or implicitly depend on the effect of the \feint and its following actions; and (3) for different Dual-Behavior models in different gaming scenarios, the threshold that divides short-term and long-term should be dynamically adjusted to enable sufficient flexibility in strategy making.

For \textit{Dynamic Short-Long-Term}, we use the time-step length of a Dual-Behavior Model $t_s$ as the short-term planning threshold. The short-term (the Dual-Behavior) starts at time step $t_0$ with actions of a \feint behavior $\{a_{t_0}^i,...,a_{t_0+t_f}^i\}$ and actions of a high-reward behavior $\{a_{t_0+t_f+1}^i,...,a_{t_0+t_s}^i\}$ sampled from \feint policy $\pi'$ ($t_f$ denotes the \feint behavior length). We use two different set of scheduler weights on Feint behaviors $\alpha_{Feint} = \{\alpha_{t_0}, ...,\alpha_{t_0+t_f}\}$ and the following attack behavior $\alpha_{attack} = \{\alpha_{t_sf}, ...,\alpha_{t_0+t_s}\}$. $\alpha_{Feint}$ are small weights as regularizers since the \feint behaviors themselves do not intend to gain rewards, while $\alpha_{attack}$ are large weights to emphasize the importance of the intended behaviors that \feint lead to. Thus, we formalize the short-time reward as
\begin{equation}
\label{equa:short-term-reward}
    Rew_{short}(\pi_i^{'}, t_0, t_f, t_s, \bm{\alpha}) = \sum_{t=t_0}^{t=t_0+t_f}\alpha_{{Feint}_t} R^i(s_t, a_t^i, a_t^{-i}) + \sum_{t=t_sf}^{t=t_0+t_s}\alpha_{{attack}_t} R^i(s_t, a_t^i, a_t^{-i})
\end{equation}

, since the purpose of \feint policy $\pi_i'$ is to actively find effective combinations of \feint behaviors and high-reward behaviors in Dual-Behavior Models that can benefit in a short-term period. We then consider long-term planning after the short-term planning threshold $t_s$: we use a set of discount factor $\beta = \{\beta_{t_0+t_s+1},...,\beta_T\}$ on the long-term average reward calculation (proposed by \cite{ICLR22/Active-Markov-Game}), to distinguish these reward from short-term rewards:
\begin{equation}
\label{equa:long-term-reward}
    Rew_{long}(\pi_i^{'}, t_0, t_s, T, \bm{\beta}) = \frac{1}{T} \sum_{t=t_0+t_s+1}^T \beta_tR^i(s_t, a_t^i, a_t^{-i})
\end{equation}
where $T$ denotes the end time of the long-term planning threshold. 

Finally, we put them together to formalize the \textit{Short-Long-Term} reward calculation mechanism, when an agent $i$ plans to perform a \feint action at time $t_0$ with a short-term planning threshold $t_s$ and the end time $T$ as:
\begin{align}
\label{Equation:short-long-term}
    Rew_{temporal}(\pi_i^{'}, t_0, t_f, t_s, T, \bm{\alpha}, \bm{\beta}) = & \ \lambda_{short}Rew_{short}(\pi_i^{'}, t_0, t_f, t_s, \bm{\alpha}) \nonumber \\
    & + \lambda_{long}Rew_{long}(\pi_i^{'}, t_0, t_s, T, \bm{\beta})
\end{align}

where $\lambda_{short}$ and $\lambda_{long}$ are weights for dynamically balancing the weight of short-term and long-term rewards for different gaming scenarios. $\lambda_{short}$ and $\lambda_{long}$ are initially set as $0.67$ and $0.33$ and are adjusted to achieve better performance with the iterations of training.
\vspace{-8pt}

\subsubsection{Spatial Dimension: Influence Range}
\label{subsubsec:strategy-spatial}
\vspace{-8pt}

The spatial advantage of \feint behaviors refers to deceiving the opponents (i.e., deviating the opponents' actions from their policies while exploring new possible advantage game states). In a Multi-Player Game (i.e. usually more than two players), the strict one-to-one relationship between two agents is not realistic, since an agent can impact both the target agent and other agents. Therefore, the influences on all other agents shall maintain different levels \cite{NIPS21/BD-RD}. Therefore, our work includes the spatial dimension of \feint impacts by fusing spatial distributions (i.e., joint state-action space distribution). The key idea of this design is to model the influence range of Feint behaviors as the divergence of occupancy measures during policy learning. More specifically, we incorporate Behavioral Diversity from \cite{NIPS21/BD-RD}, to mathematically calculate and maximize the diversity gain of \feint behaviors on the influence range. 

We follow the occupancy measure introduced in~\cite{NIPS21/BD-RD}, the distribution of state-action pairs $\rho_{\bm{\pi}}(s, \bm{a}) = \rho_{\bm{\pi}}(s) \bm{\pi}(\bm{a} \mid s)$, to measure a joint policy $\bm{\pi} = (\pi_i, \pi_{-i})$. $\rho_{\bm{\pi}}(s)$ can be calculated by the normalized-weighted sum of game state visit probabilities for the joint policy $\bm{\pi} = (\pi_i, \pi_{-i})$ of all agents. Game state $s$ can be parameterized by a set of physical properties. We demonstrate a set of commonly used parameters in boxing games \cite{SIGGRAPH21/Two-Player-Game}: the relative positions $p(i, -i)$, relative moving orientations $o(i, -i)$, the linear velocities $l\_{vel}(i, -i)$, and angular velocities $a\_{vel}(i, -i)$. $s$ can thus be composed in a vector $s = (p(i, -1), o(i, -i), l\_{vel}(i, -i), a\_{vel}(i, -i))$. The spatial domain influence of \feint policy can be naturally represented by the occupancy measure. When players apply \feint behaviors to deceive opponents, the resulting spatial impact is to exploit new possibilities to achieve more advantageous state transitions from the current state. When a \feint policy $\pi_i'$ is added, we aim to maximize the effective influence range under the influence distribution of \feint. Specifically, we maximize the divergence of the new distribution of joint policy $\bm{\pi}' = (\pi_i', \pi_{-i})$ introduced by \feint from the distribution of the old one $\bm{\pi} = (\pi_i, \pi_{-i})$ without \feint. By using Behavior Diversity~\cite{NIPS21/BD-RD}, such a maximization problem at state $s$ can be formalized as:
\begin{equation}
    max_{\substack{\pi_i'}}Rew_{spatial}(\pi_i^{'}, \pi_{-i}, s)=D_f(\rho_{\pi_i', \pi_{{-i}}}(s) \mid\mid \rho_{\pi_i, \pi_{{-i}}}(s))
\end{equation}
where we use the divergence as the reward $Rew_{spatial}$ and the general $f$-divergence is used to measure the discrepancy of two distributions. \cite{NIPS21/BD-RD} introduces multiple ways to approximately calculate such divergence for policy models represented by neural networks.

\vspace{-8pt}

\subsection{Collective Impacts: Influence Degree}
\label{subsubsec:strategy-collective}
\vspace{-8pt}

Solely relying on the Temporal Dimension and Spatial Dimension overlooks the interactions between them, and these two dimensions are expected to have mutual influences for realistic modeling \cite{NIPS21/BD-RD}. Therefore, we consider the influence degree for the collective impacts.

We formulate it for a \feint policy $\pi_i'$ in Multi-Player Games which performs a full \feint cycle (i.e., a complete Dual-Behavior Model and long-term actions) that starts at $t_0$ and ends at $T$ as:
\begin{align}
\label{Equation:collective-impact}
    Rew_{collective}(\pi_i^{'}, \pi_{-i}) = & \ \mu_1 \sum_{\pi \in \{\pi_i^{'}, \pi_{-i}\}} 
    Rew_{temporal}(\pi, t_0, t_f, t_s, T, \bm{\alpha}, \bm{\beta}) \nonumber \\
    & + \mu_2 \sum_{s=s_0}^{s_T} Rew_{spatial}(\pi_i^{'}, \pi_{-i}, s)
\end{align}
where temporal impacts $Rew_{temporal}$ (Section~\ref{subsubsec:strategy-temporal}) for agent $i$ are are aggregated on spatial domain considering all agents and spatial impacts $Rew_{spatial}$ (Section~\ref{subsubsec:strategy-spatial}) are aggregated on the temporal domain for all the states in the time period. $\mu_1$ and $\mu_2$ denote the weights of aggregated temporal impacts and spatial impacts respectively, enabling flexible adaption to different gaming scenarios. They are initially set as $0.5$.

In addition to the collective impacts of \feint itself in terms of temporal domain and spatial domain, our formalized impacts of \feint can also result in response diversity of opponents, since different related opponents (spatial domain) at different time steps (temporal domain) can have diverse response. Such Response Diversity can be used as a reward factor that makes the final reward calculation more comprehensive \cite{ICML21/BD, NIPS21/BD-RD}. Thus, to incorporate the Response Diversity together with our final reward calculation model, we refer to \cite{NIPS21/BD-RD} to characterize the diversity gain incurred by our collective impacts formalization. For an agent who maintains a pool of policy $\mathbb{P}_{i} = \{\pi_i^1, …, \pi_i^M\}$ while opponents maintain a pool of policy $\mathbb{P}_{-i} = \{\pi_{-i}^1, …, \pi_{-i}^N\}$, an empirical payoff matrix $A_{\mathbb{P}_{i} \times \mathbb{P}_{-i}}$ can be calculated element-wise using the reward function $Rew_{collective}(\pi_i^k, \pi_{-i}^j)$ for $(k,j)$ entry. Thus the Response Diversity gain for a \feint policy $\pi^{M+1}$ can be measured by follows:
\begin{equation}
\label{Equation:collective-diversity}
    Rew_{collective-diversity}(\pi_i^{M+1}) = D(\bm{a}_{M+1}\mid\mid A_{\mathbb{P}_{i} \times \mathbb{P}_{-i}})
\end{equation}
\begin{equation}
\label{Equation:new-action-reward}
    \bm{a}_{M+1}^T:=(Rew_{collective}(\pi_i^{M+1}, \pi_{-i}^j))_{j=1}^N.
\end{equation}
where $D(a_{M+1}\mid\mid A_{\mathbb{P}_{i} \times \mathbb{P}_{-i}})$ represents the diversity gain of the \feint action on current policy space. We follow the method in \cite{NIPS21/BD-RD} for the quantification of diversity gain, which uses a practical and differential lower bound for feasible computation.

\vspace{-12pt}

\section{Proof-of-concept Implementation}
\label{sec:implementation}
\vspace{-8pt}

To provide a unified implementation scheme of \feint into most MARL frameworks, we choose to implement on the training iteration level and avoid changing the MARL models themselves. We create an additional policy model (e.g., MADDPG~\cite{NIPS17/MADDPG}, MASAC~\cite{ICML18/SAC, ICML19/MAAC}, MAD3PG~\cite{ICLR18/D4PG,MDPI21/MAD3PG}, MATD3~\cite{NIPS19/MATD3}, etc.) for each agent as the \feint policy, which works together with the regular policy models for agents but is trained and inferenced differently. 

\begin{figure}[h]
    \centering
    \includegraphics[width=\linewidth]{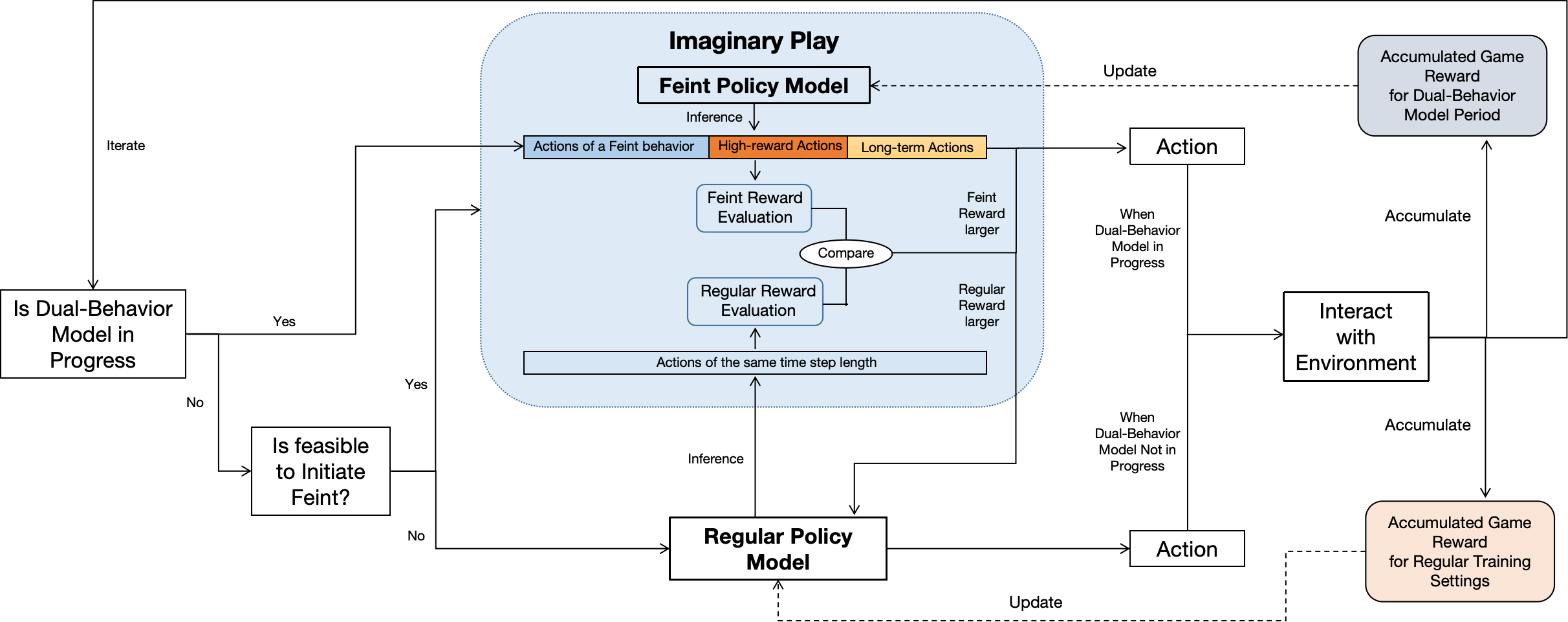}
    \caption{Illustration of \feint behavior implementation in game iterations}
    \label{fig:implementation}
\end{figure}

Figure~\ref{fig:implementation} illustrates the full process of our implementation of \feint in game iterations. We implement the \feint behavior generation in an imaginary play module in training iterations (i.e., game steps). The imaginary play module decides whether an agent should initiate a \feint behavior, composes a Dual-Behavior Model using Palindrom-directed templates, and utilizes the \feint reward calculation to evaluate the quality of the generated action sequence in the Dual-Behavior model. 

In our experimental settings, \feint behavior templates can be pre-computed only once before training, providing a fast lookup for composing Dual-Behavior models in gaming iterations. Algorithm~\ref{algo::feint-templates} in Appendix~\ref{appendix:implementation} illustrates the pseudo-code for pre-computing available \feint behavior templates given a set of available attack behaviors $B$. For each pair of attack behaviors $(Behavior_i, Behavior_j)$ in $B$, we check whether there are physically satisfiable states illustrated in Figure~\ref{fig:three-stage}. If any of the three conditions are satisfied, we cut out the available actions ($Avail_i$ and $Avail_j$) and store their compositions as an available template candidate. Note that the stored templates do not enforce the \feint behaviors to contain the exact same action sequences. Instead, tuples $(a_k, Avail_i, Avail_j)$ are served as keys for quick lookup and action restrictions in composing Dual-Behavior models during gaming iterations.

During gaming iterations, the imaginary play will only be activated when no Dual-Behavior Model is in progress and the current physical state $s_c$ of an agent is close to a physical state $s_r$ where it is physically realistic for the agent to perform a high-reward action $a_{target}$, while the possibility of performing $a_{target}$ is relatively low according to its regular policy model (i.e., action $a_{target}$ are highly likely to be diminished by other agents current actions). Thus, the purpose of \feint behavior is to lead the agent to a state $s_r$ where the agent can maximize the game environment reward by performing the intended high-reward action $a_{target}$ (i.e., other agents are deceived by \feint to perform other actions which cannot effectively diminish the high-reward actions performed by the agent). 

When the imaginary play is activated, available Dual-Behavior models can be generated using the last known action $a_t$ and the intended high-reward action $a_{target}$. Algorithm~\ref{algo:dual-behavior-models} in Appendix~\ref{appendix:implementation} shows the pseudo-code for composing available Dual-Behavior models with backward searches. The intended high-reward action $a_{target}$ provides constraints on the ending actions of the \feint behaviors, while the last known action $a_t$ provides constraints on the starting action of \feint behaviors. Thus, the available Dual-Behavior models can be quickly computed by doing a one-pass search from the pre-computed \feint behavior templates. The composed Dual-Behavior models can thus constrain the available action choices for the \feint policy model (set other actions' possibilities to 0) in corresponding templates and use a reflection frame to compose a (semi-)palindrome leading to the agent's physical state $s_r$.
After having available Dual-Behavior models which are composed of \feint behavior and followed by high-reward actions, the short-term reward can be calculated by Equation~\ref{equa:short-term-reward}. After this Dual-Behavior action sequence, the imaginary play would play a few steps to incorporate the long-term reward (using Equation~\ref{equa:short-term-reward}). The collective reward (Section~\ref{subsubsec:strategy-spatial}) can thus be calculated. This reward is then compared to an accumulated reward from an imaginary play using only the agent's regular policy model in the same number of time steps. If the \feint collective reward is higher, the action sequence of the dual action model will be applied in the following real-game steps. When a Dual-Behavior Model is in progress, the actions will not be sampled from the regular policy models.

In the real game steps, where all the agents' actions interact with the environment and the real game rewards are calculated, our formalization of \feint only changes the way to update the \feint policy models for agents. The \feint policy models are updated only when corresponding Dual-Behavior Models finish and are updated using the accumulated real game rewards for that period. The regular policy models are updated as usual settings (e.g., after some fixed steps - an episode).

\vspace{-12pt}

\section{Experimental Studies}

\vspace{-8pt}

\subsection{Methodology}
\label{subsec:experiment-methods}
\vspace{-8pt}

\textbf{Testbed Implementations.} Due to the lack of a general benchmark, we selectively implement two scenarios under a customized manner. They consist of a "1 vs 1" and a "3 vs 3" multi-player free fight games, based on widely-provided testbeds. We provide additional details on how our testbeds are designed and implemented in appendix~\ref{appendix:testbed}.

\textbf{Evaluation Metrics.} Our main evaluation objective is the gaming rewards. We first examine the gaming outcomes when using the MADDPG, MASAC, MATD3, and MAD3PG MARL models, by comparing the per episode gaming rewards of agents across all scenarios\footnote{Note that these rewards are the actual game rewards (the reward that returned by the gaming environment), which are not the rewards that policy models used to select actions or update parameters}. We also evaluate other metrics, and report our results in appendix~\ref{appendix:additional-results}.

\vspace{-8pt}

\subsection{Major Results}
\label{subsec:experiment-results}
\vspace{-8pt}

\begin{figure}[h]
    \parbox{.25\linewidth}{\includegraphics[width=\linewidth]{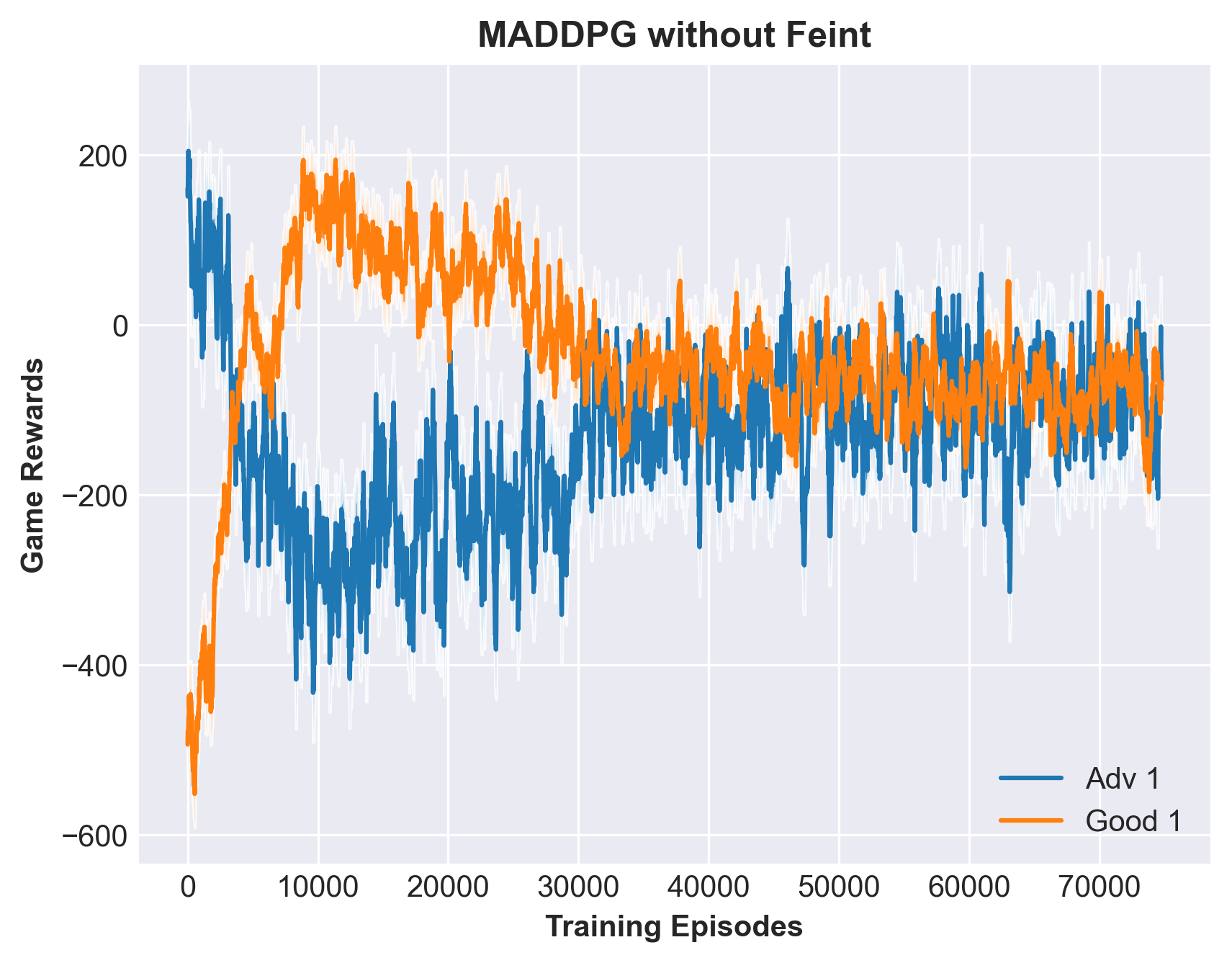}}\hfill
    \parbox{.25\linewidth}{\includegraphics[width=\linewidth]{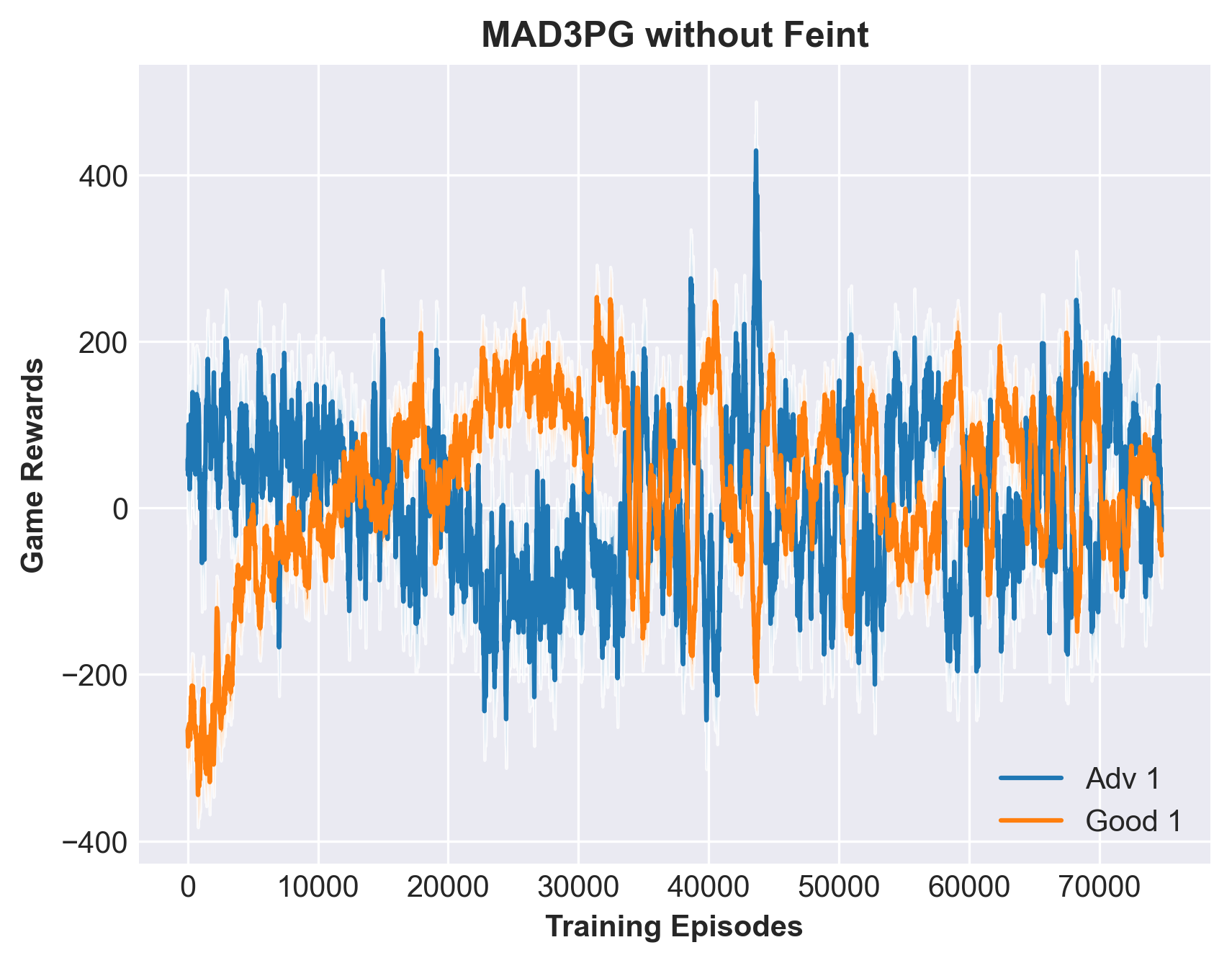}}\hfill
    \parbox{.25\linewidth}{\includegraphics[width=\linewidth]{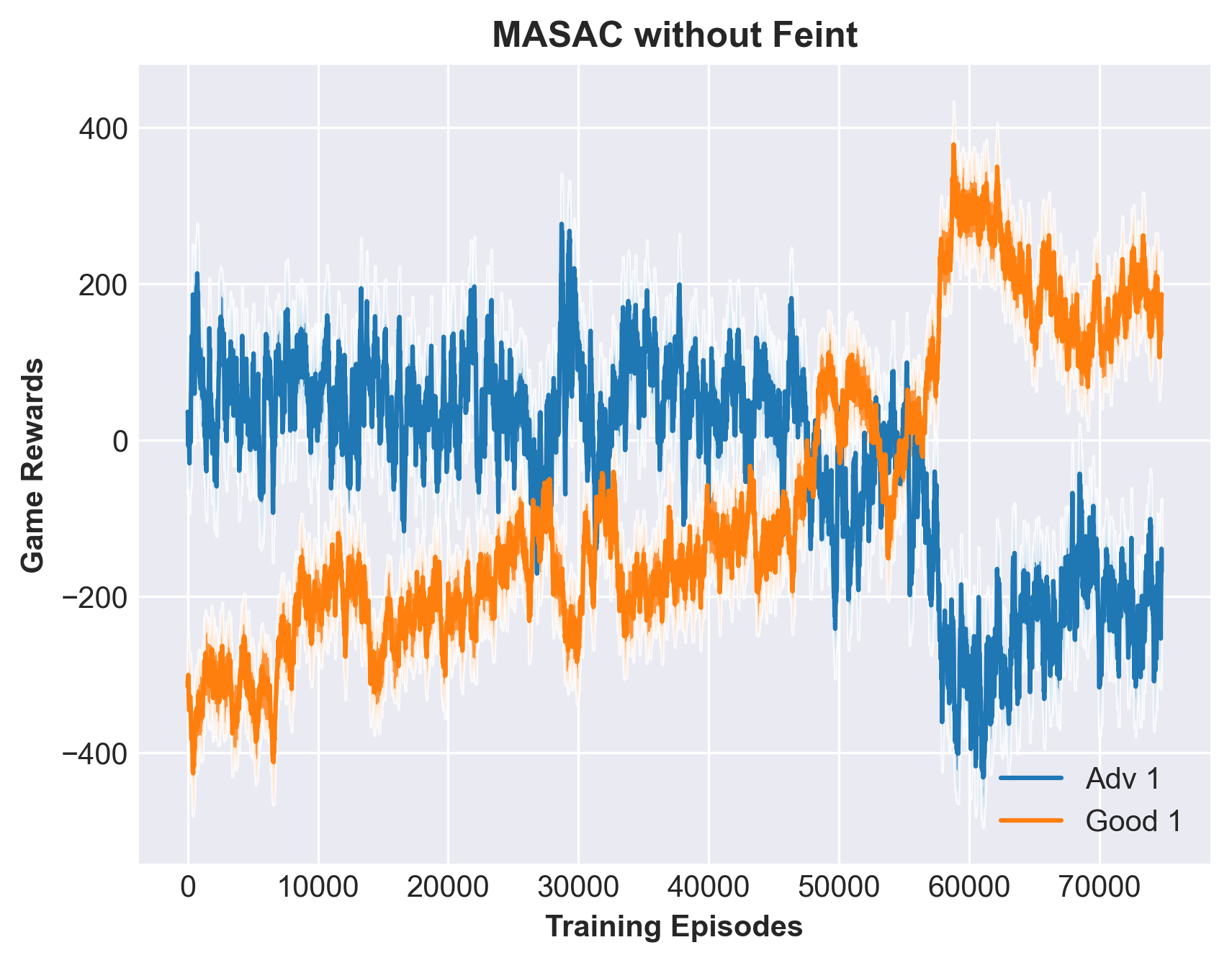}}\hfill
    \parbox{.25\linewidth}{\includegraphics[width=\linewidth]{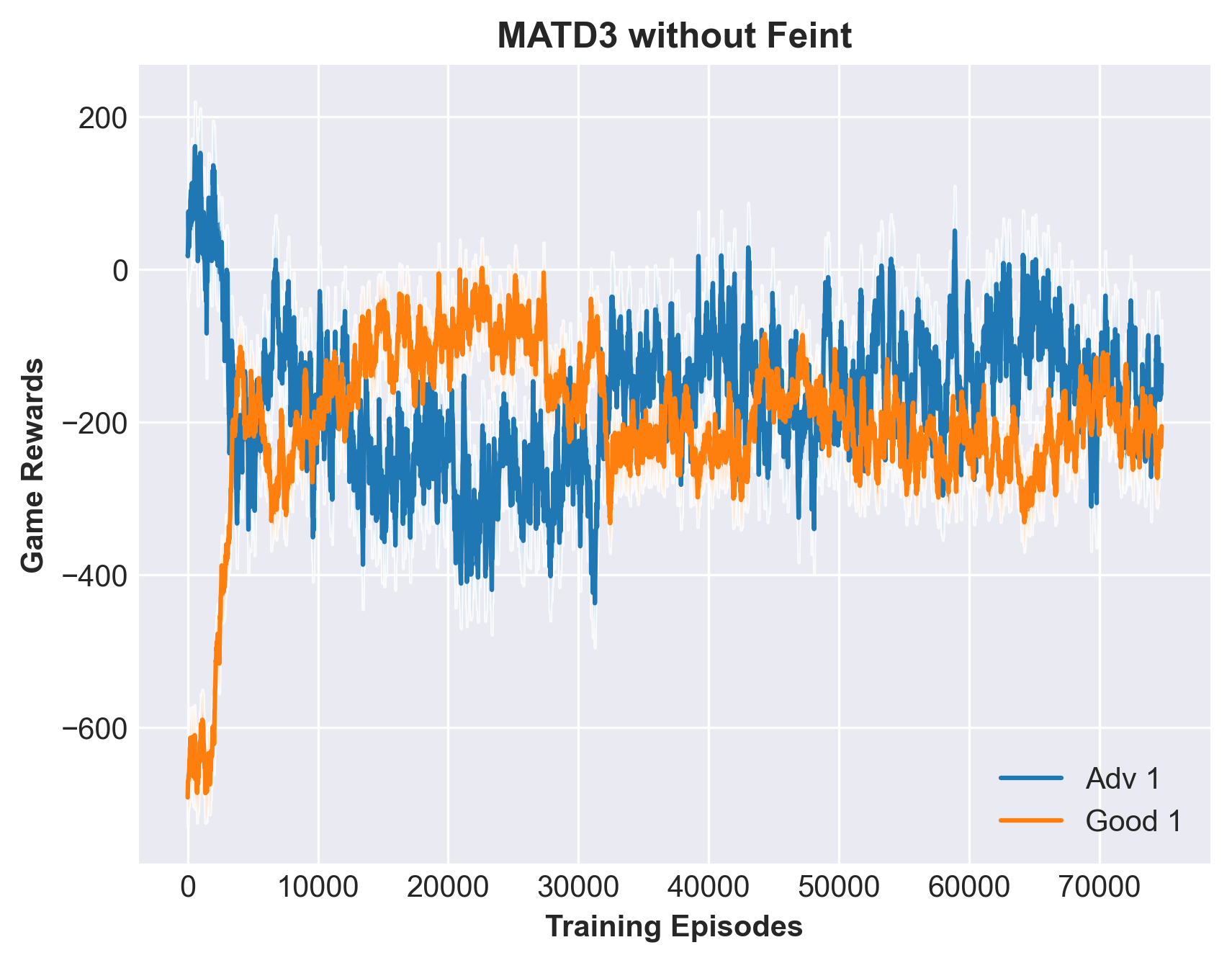}}\par
    \parbox{.25\linewidth}{\includegraphics[width=\linewidth]{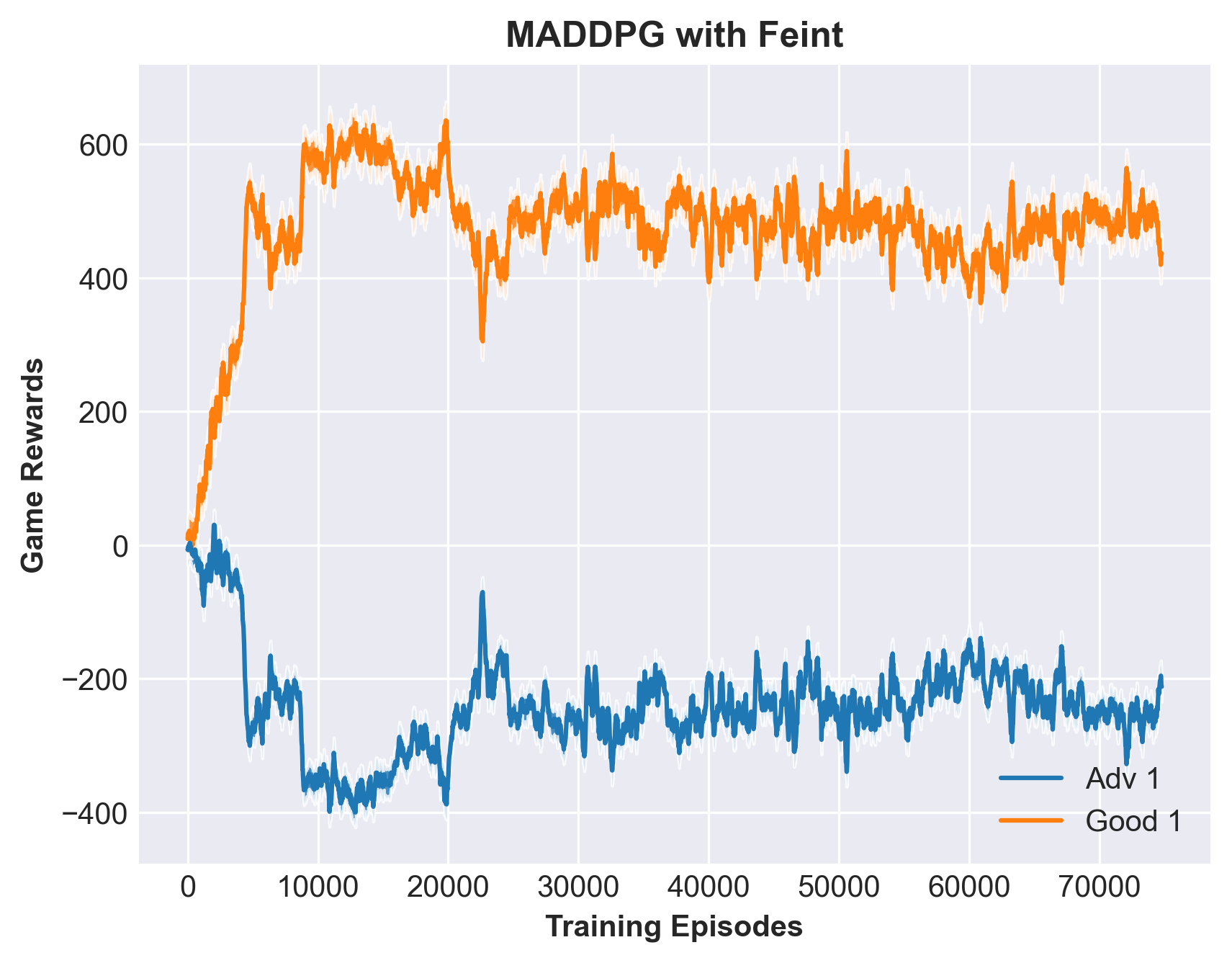}}\hfill
    \parbox{.25\linewidth}{\includegraphics[width=\linewidth]{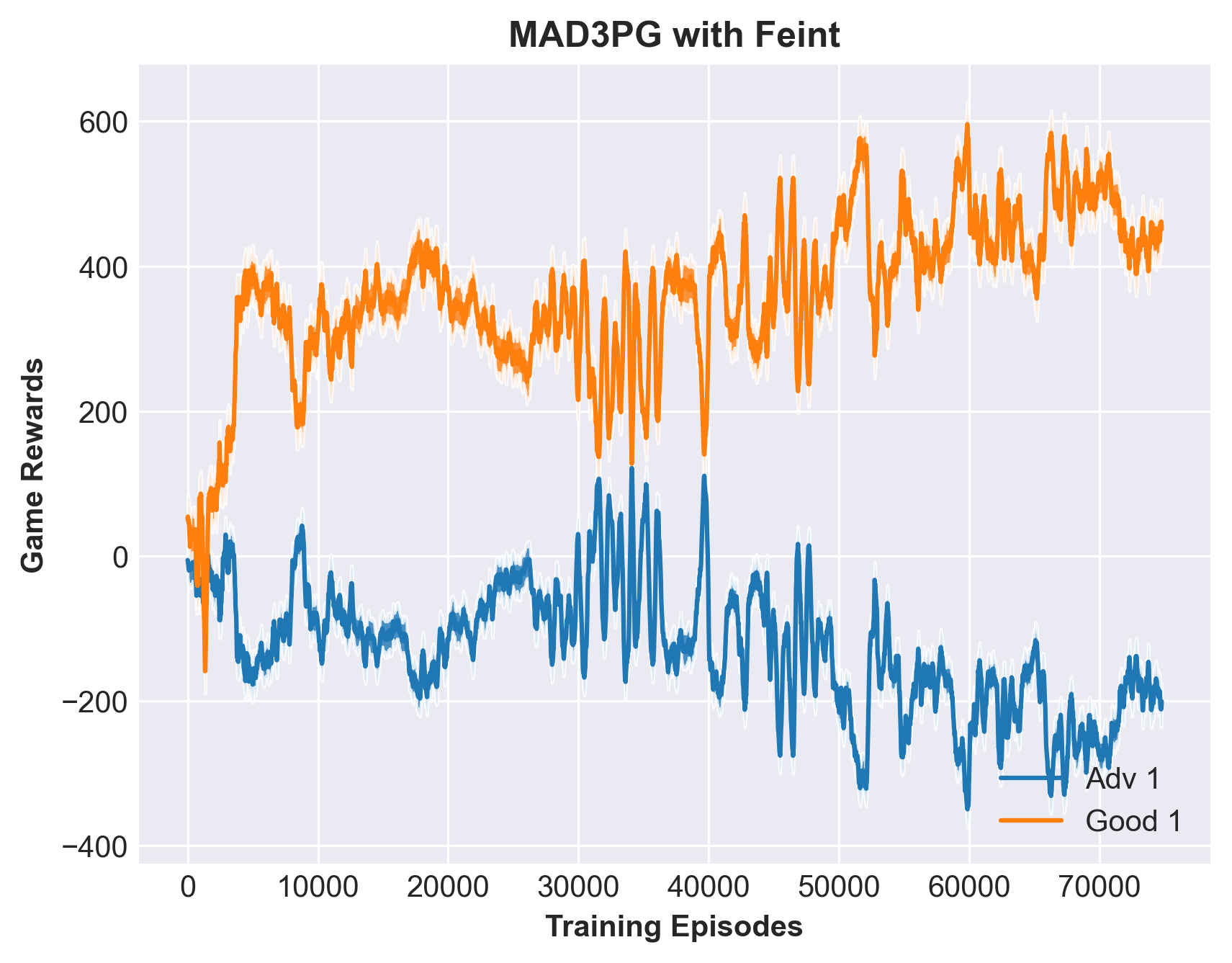}}\hfill
    \parbox{.25\linewidth}{\includegraphics[width=\linewidth]{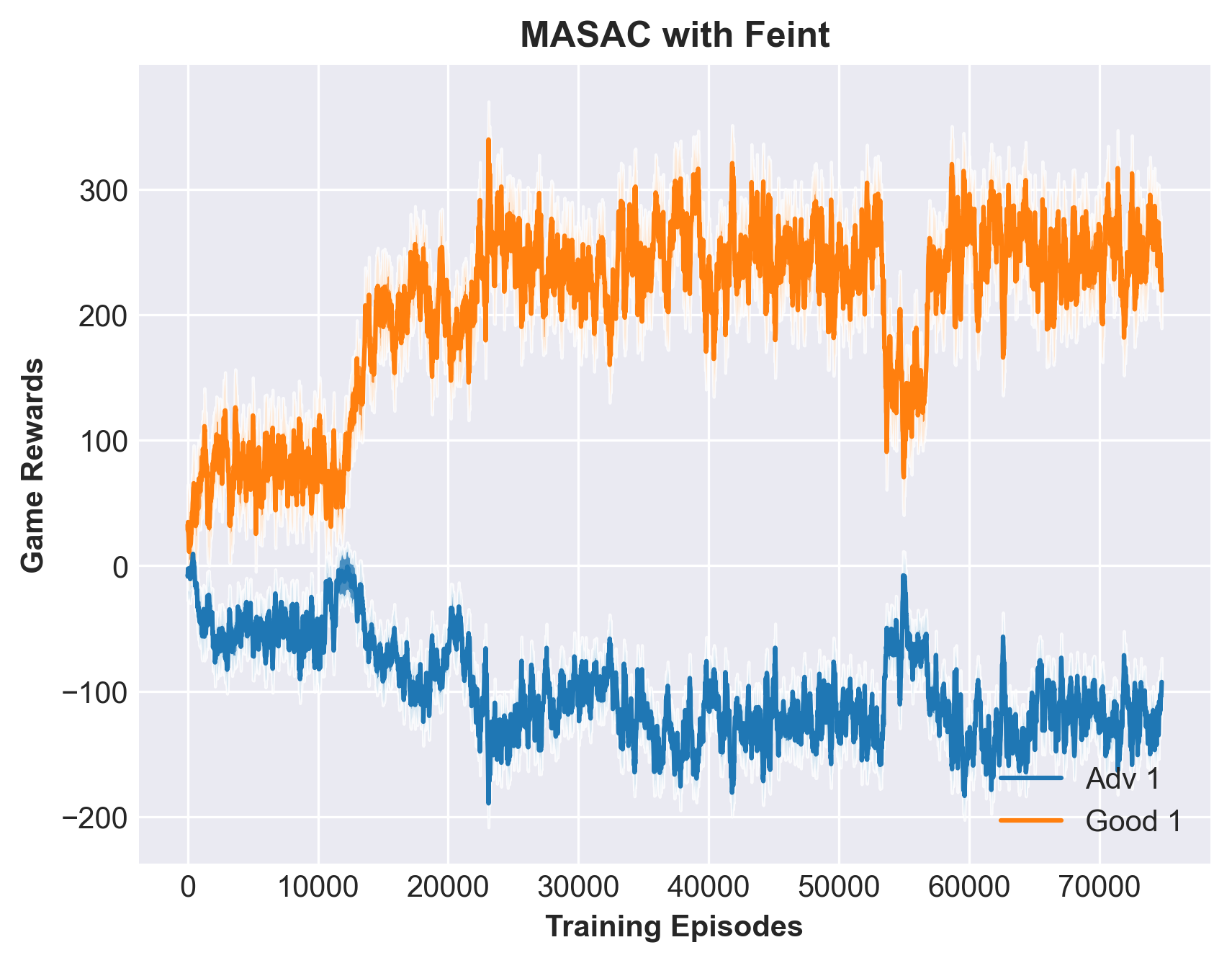}}\hfill
    \parbox{.25\linewidth}{\includegraphics[width=\linewidth]{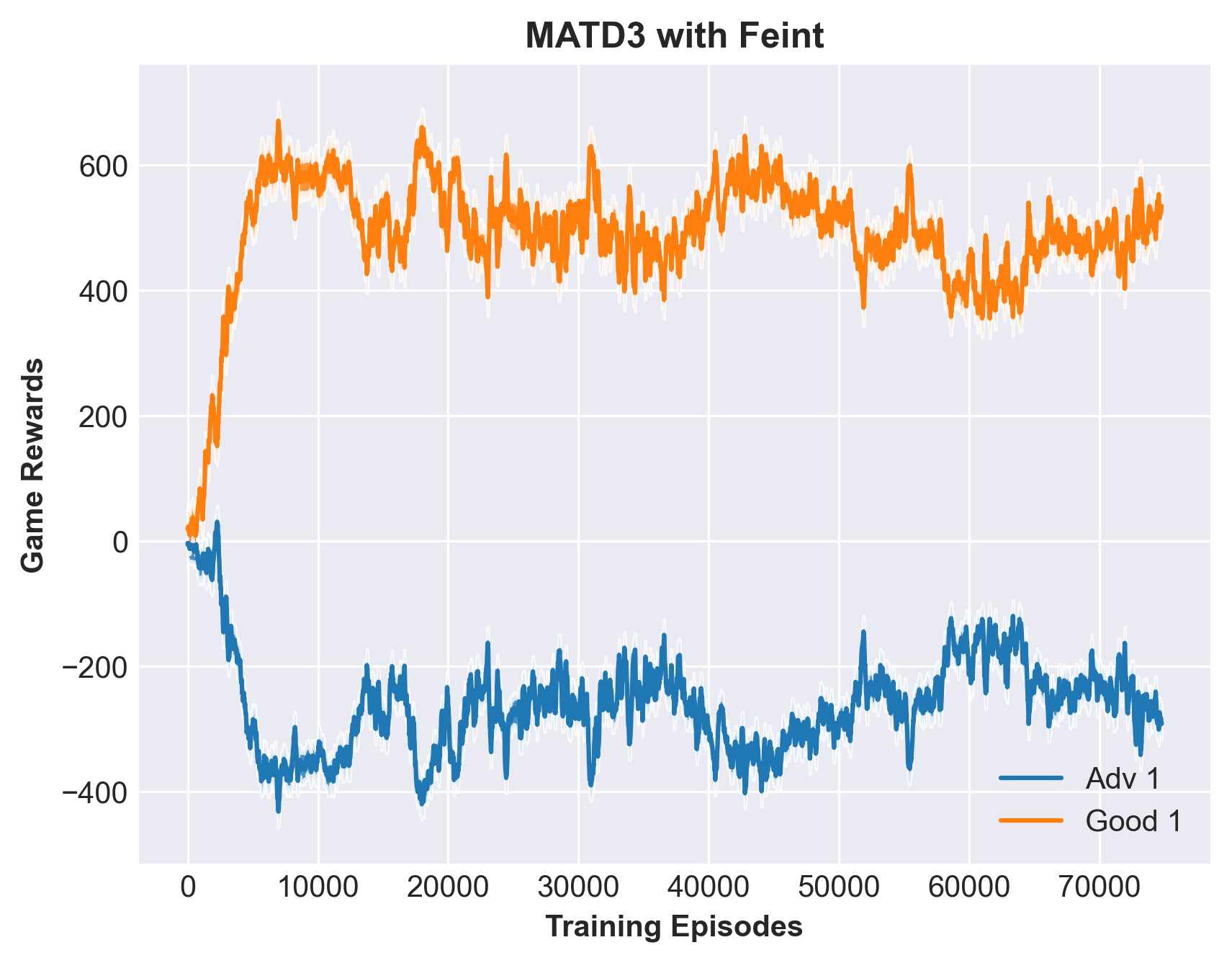}}\par
    \caption{Comparison of Game Reward when using \feint and not using \feint in a 1 VS 1 scenario.}
    \label{fig:1_vs_1}
\end{figure}

Figure~\ref{fig:1_vs_1} shows the game reward comparisons of using \feint behaviors or not in the Two-Player scenario (Section~\ref{subsec:experiment-methods}) for 4 MARL models. The first row shows the baseline results where all agents are trained normally, while the second row shows the results where the player labeled with "Good" incorporates \feint behaviors. In most of the baseline results (e.g., using MADDPG, MAD3PG, and MATD3), the two players' rewards tend to progress to a similar level when after enough training iterations. For MASAC, the "Good" player seems to gain higher rewards than its opponents when the training iterations are large, but the advantage is not stable and such a phenomenon can likely be the instability of the MASAC algorithm itself. For all the results where \feint behaviors are incorporated, we can see a significant advantage gain for the "Good" player. Thus, our formalization of incorporating \feint behaviors can effectively improve the actual game rewards in two-player combating scenarios.

\begin{figure}[h]
    \parbox{.25\linewidth}{\includegraphics[width=\linewidth]{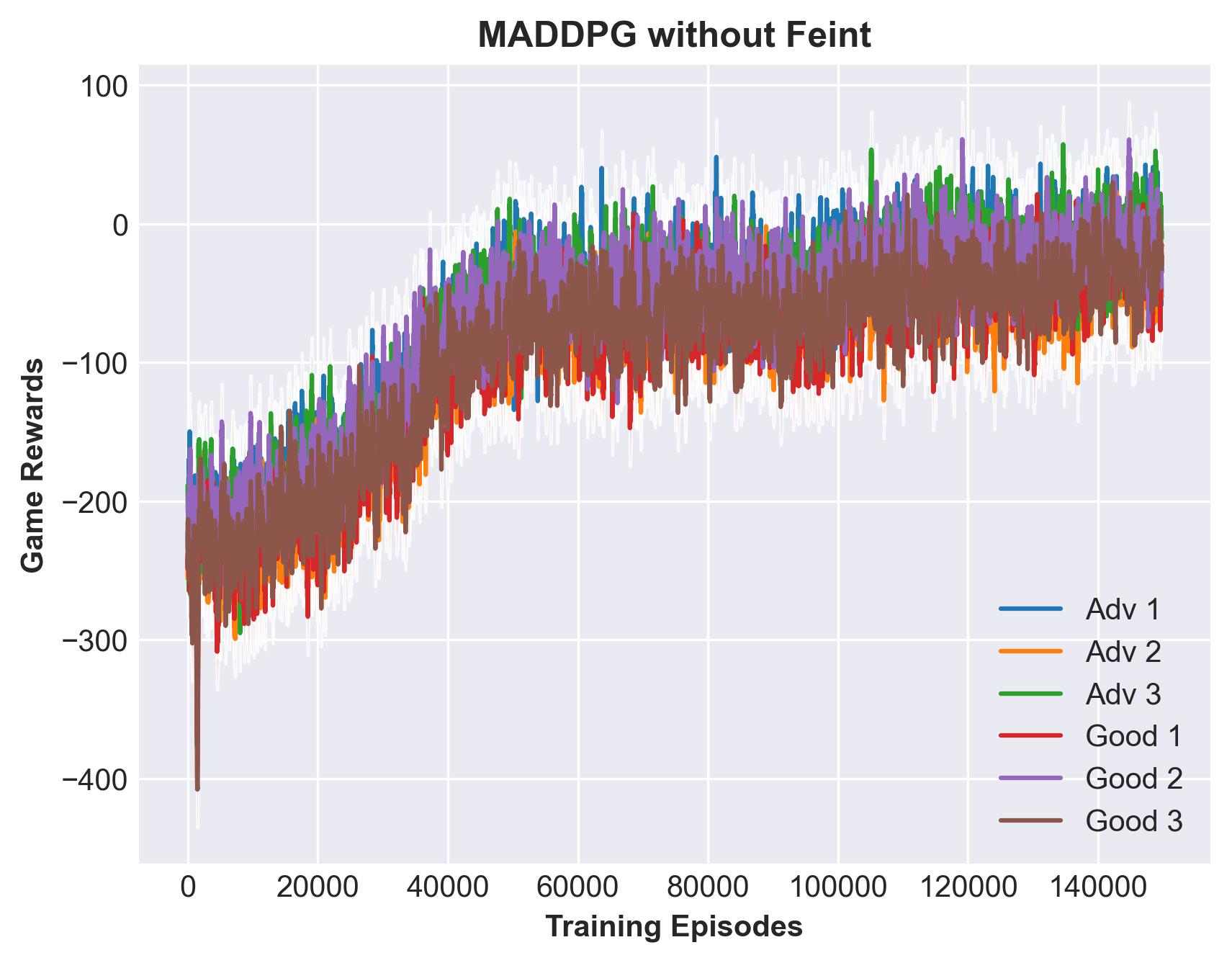}}\hfill
    \parbox{.25\linewidth}{\includegraphics[width=\linewidth]{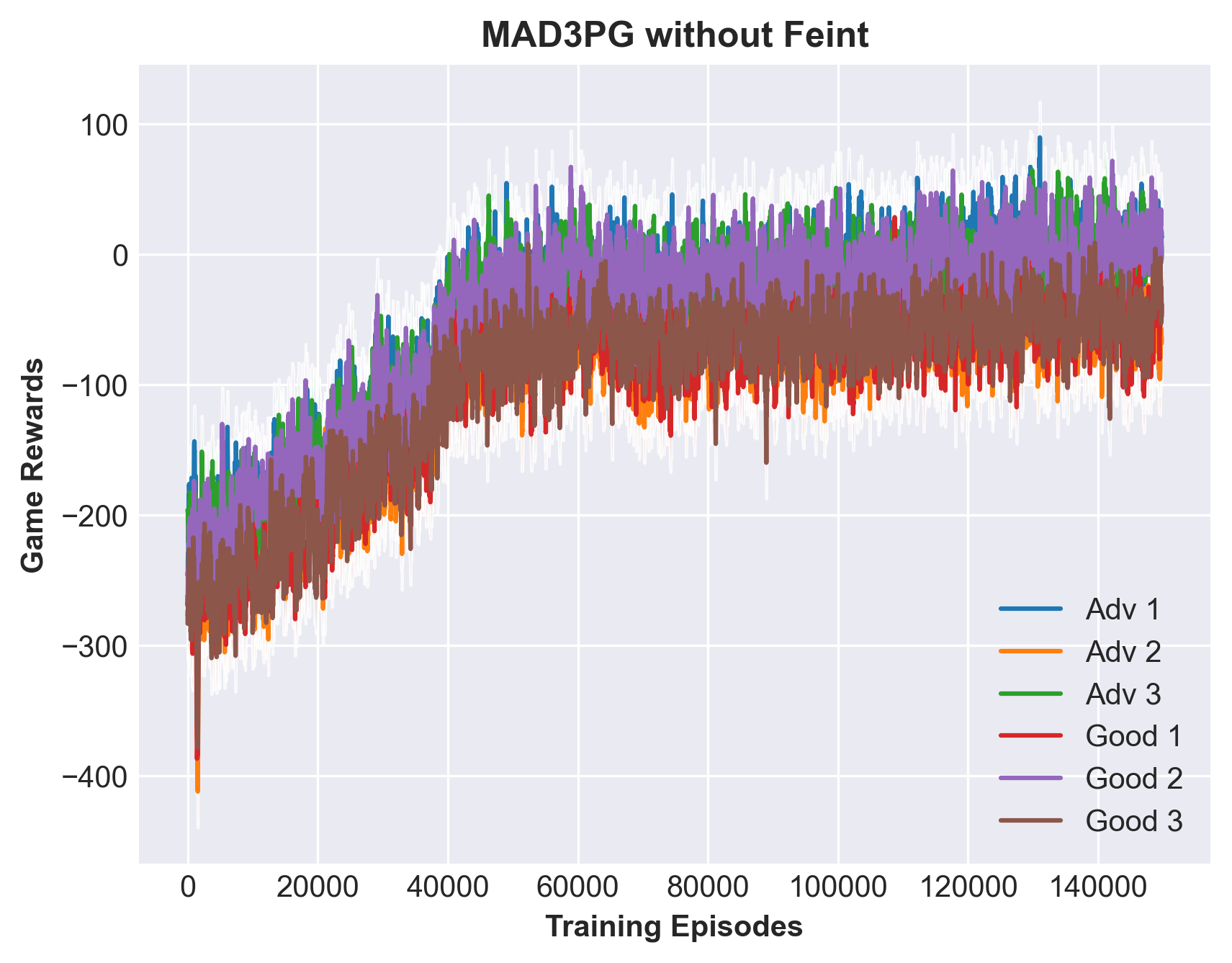}}\hfill
    \parbox{.25\linewidth}{\includegraphics[width=\linewidth]{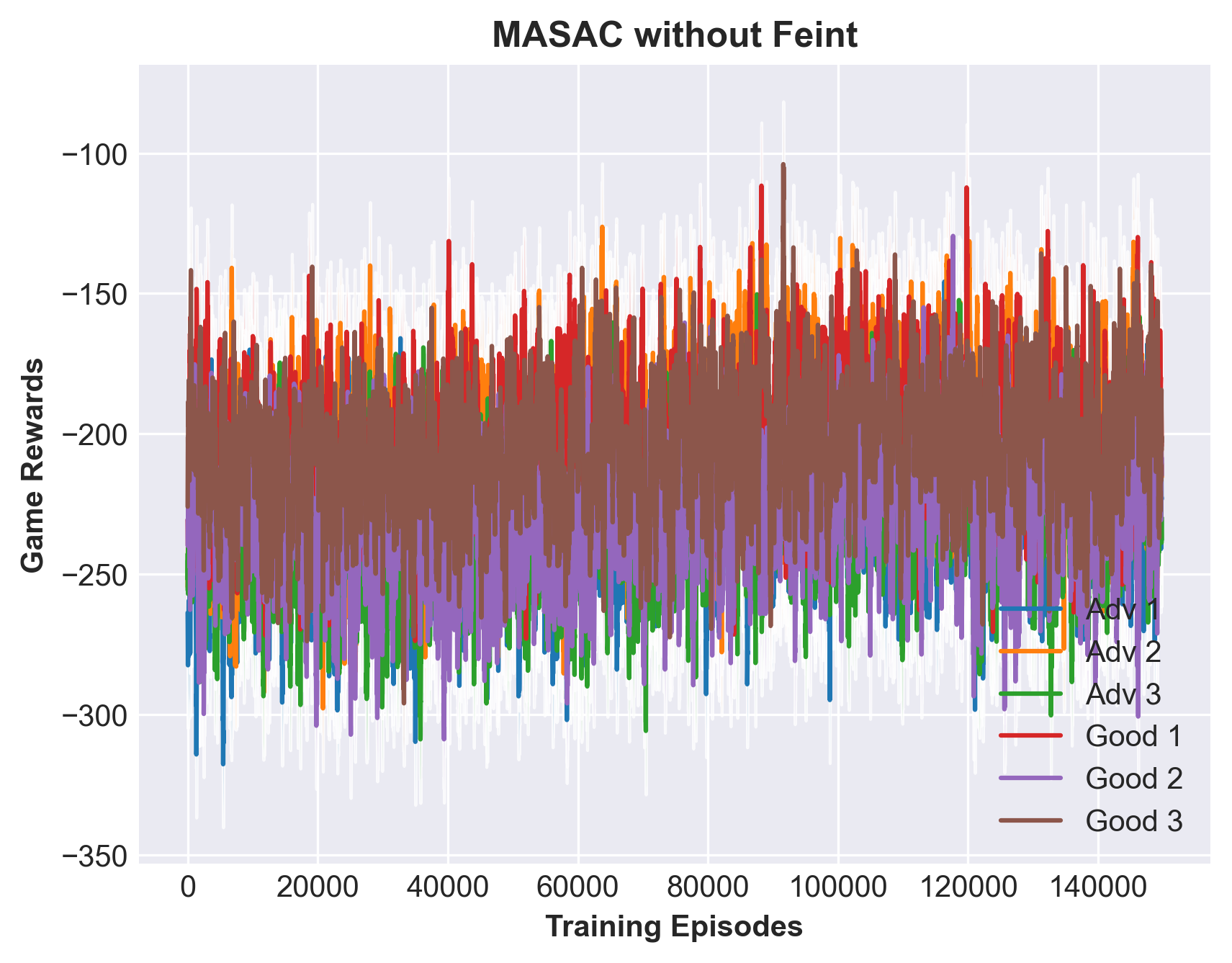}}\hfill
    \parbox{.25\linewidth}{\includegraphics[width=\linewidth]{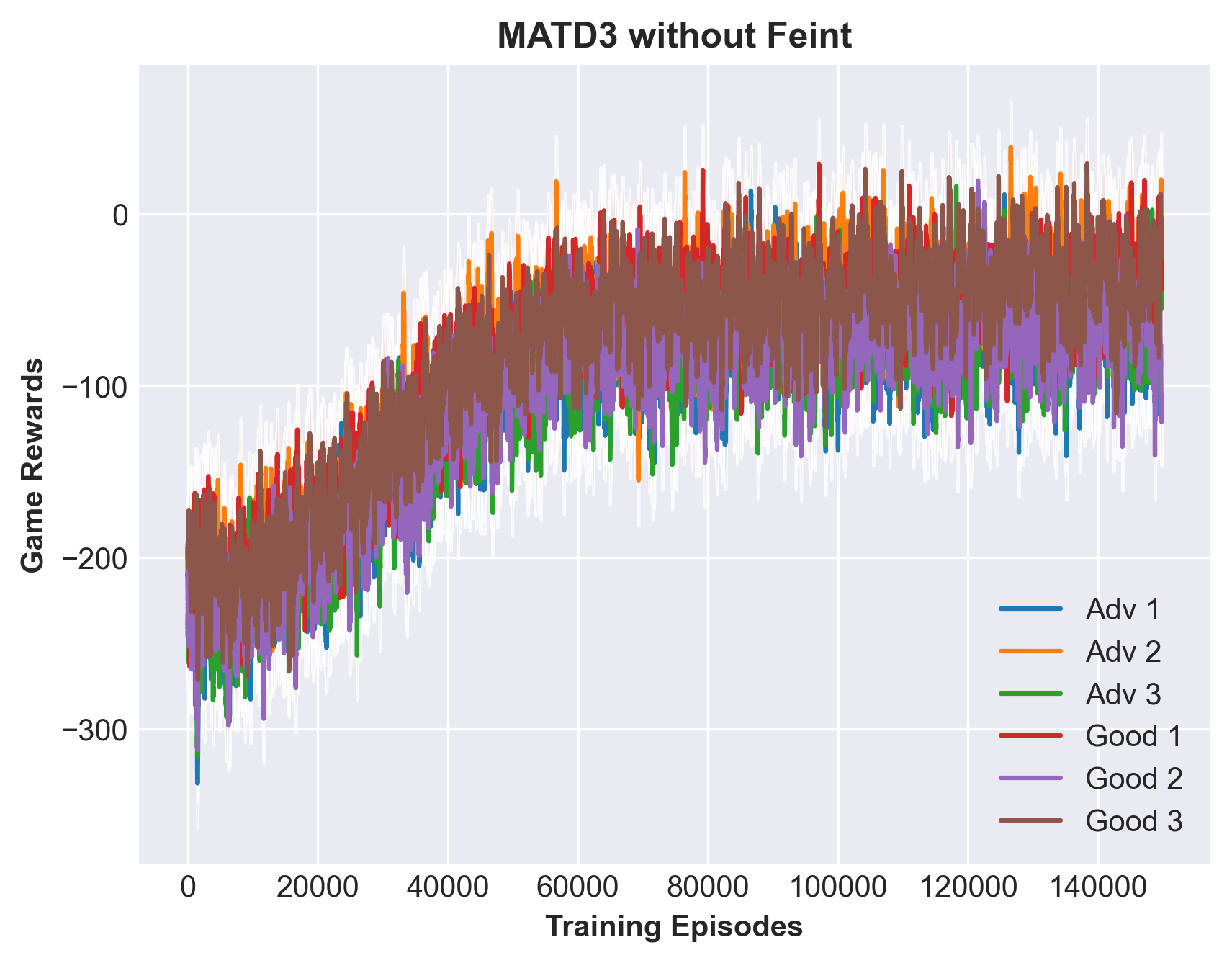}}\par
    \parbox{.25\linewidth}{\includegraphics[width=\linewidth]{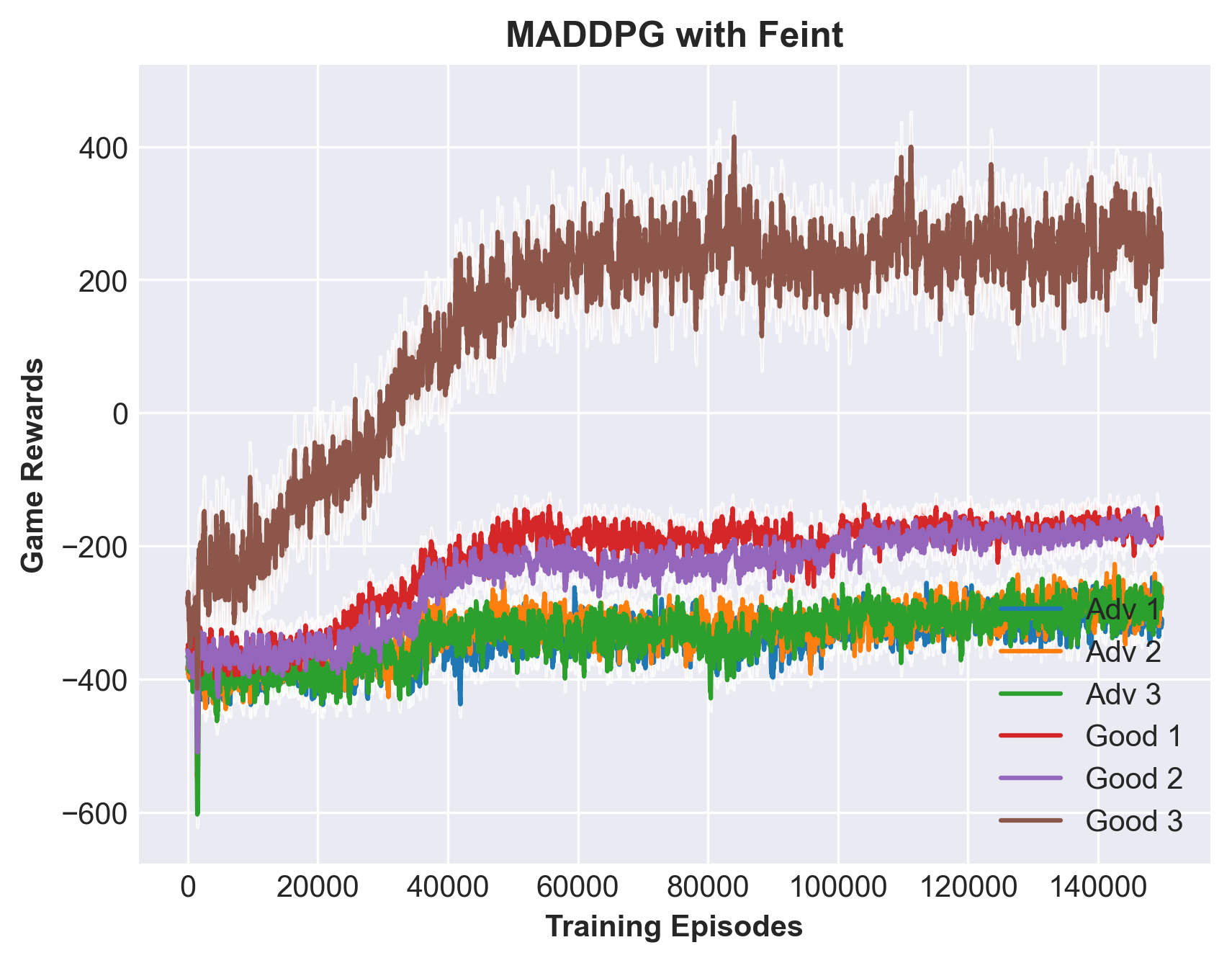}}\hfill
    \parbox{.25\linewidth}{\includegraphics[width=\linewidth]{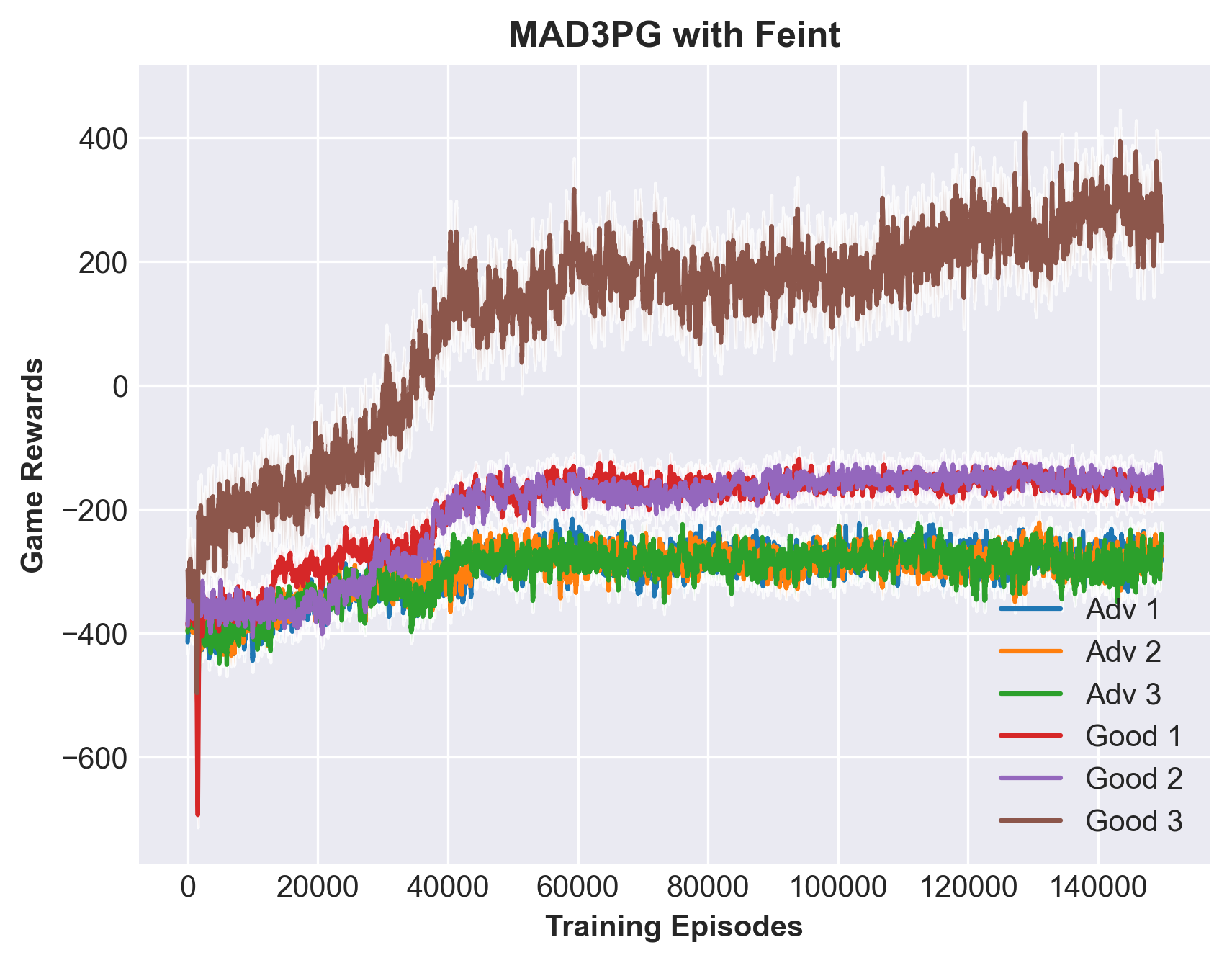}}\hfill
    \parbox{.25\linewidth}{\includegraphics[width=\linewidth]{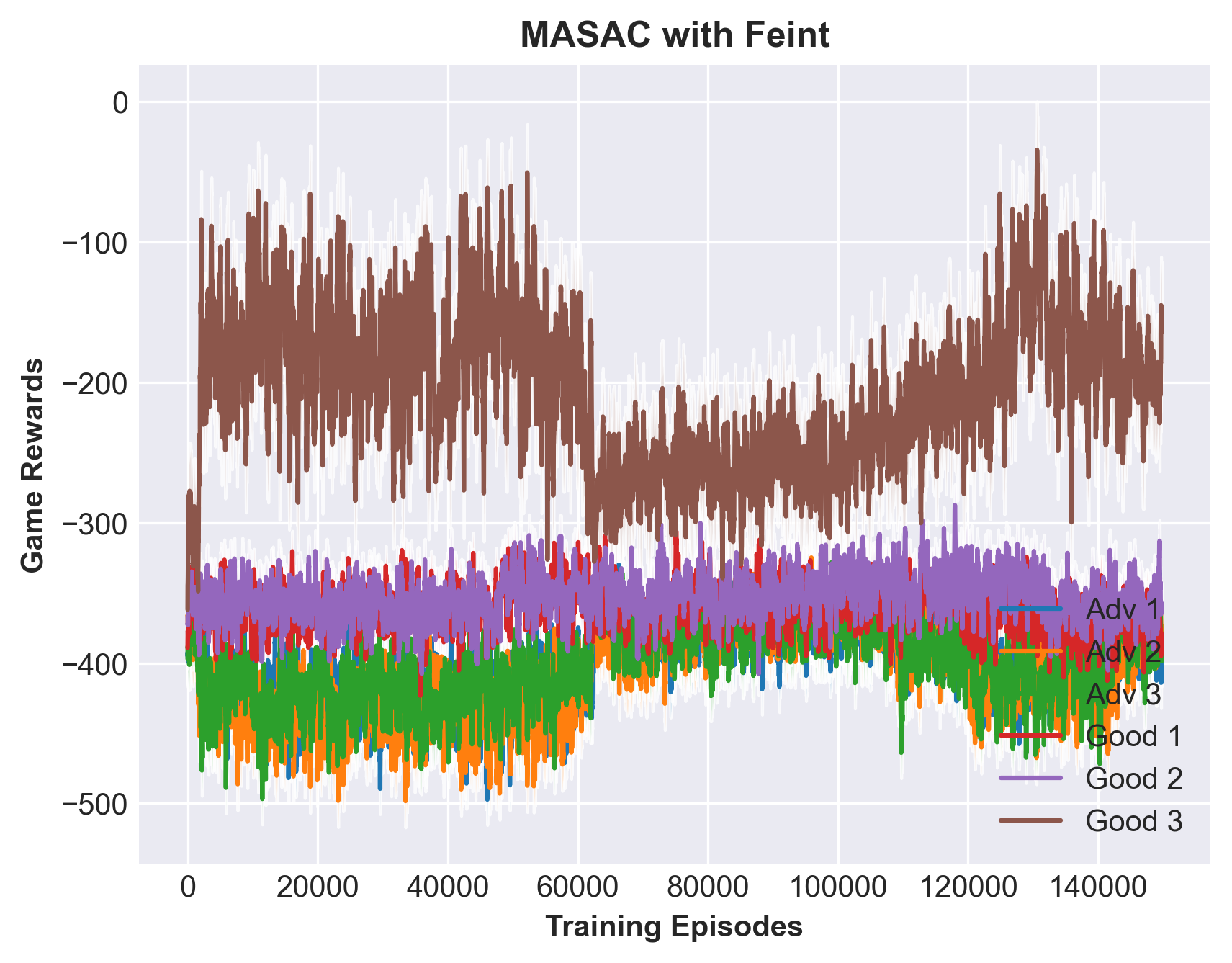}}\hfill
    \parbox{.25\linewidth}{\includegraphics[width=\linewidth]{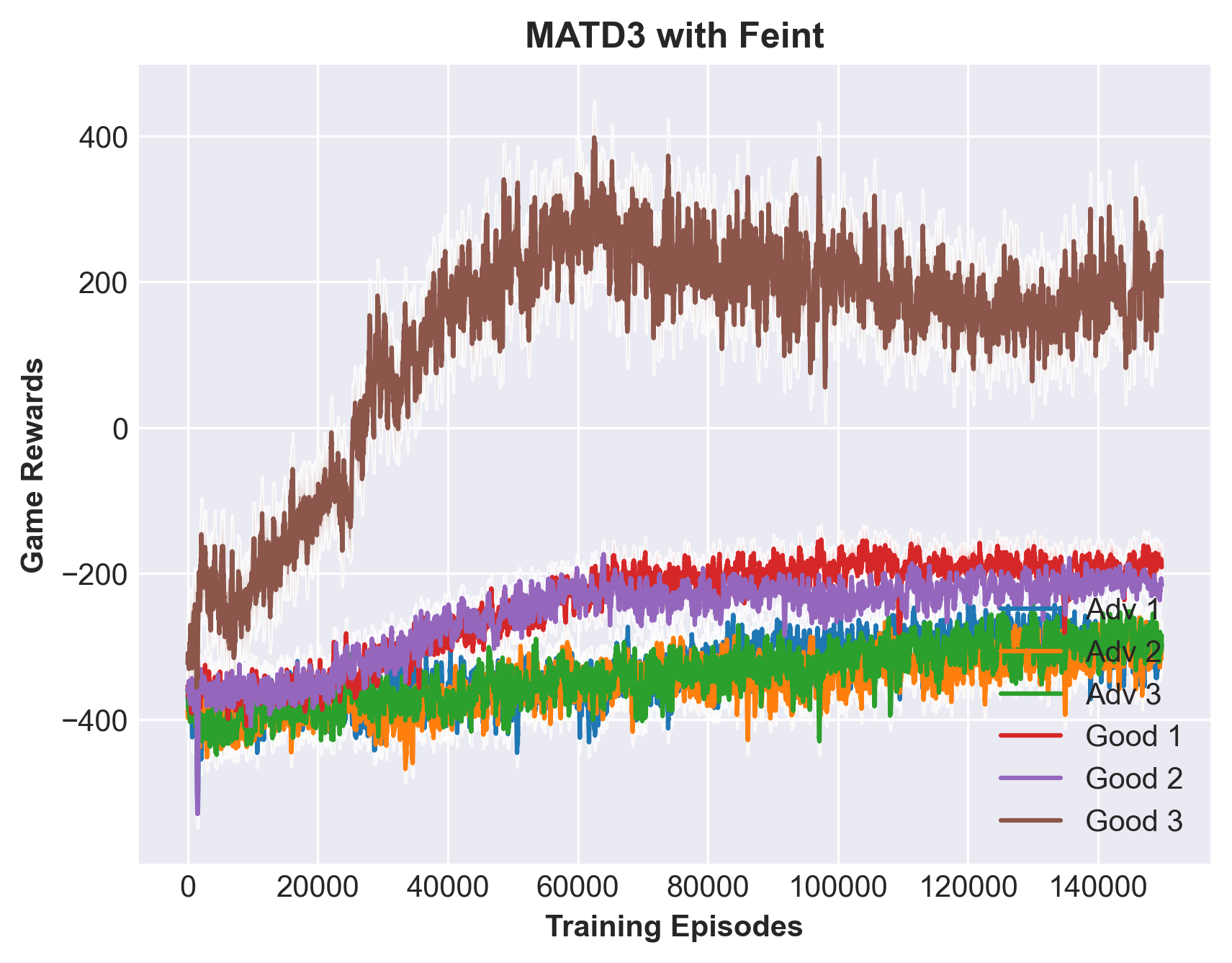}}\par
    \caption{Comparison of Game Reward when using \feint and not using \feint in a 3 VS 3 scenario.}
    \label{fig:3_vs_3}
\end{figure}

To further evaluate the effectiveness of our formalization of \feint behaviors in multi-player scenarios, Figure~\ref{fig:3_vs_3} shows the game reward comparisons in Six-Player scenario (Section~\ref{subsec:experiment-methods}) for 4 MARL models. The first row shows the baseline results while the second row shows the results where the player labeled with "Good 3" incorporates \feint behaviors. In all baseline results, all 6 players seem to achieve similar levels of rewards after enough training iterations. In comparison, in all results where the "Good 3" player incorporates \feint, it gains significantly more rewards than the opponents as well as its teammates. This result shows that our formalization of \feint can not only gain higher rewards towards the direct opponents, but also gain advantages among teammates who do not incorporate \feint. Another interesting observation is that there are no more symmetric patterns in the players' rewards, showing that the gaming interactions in multi-player scenarios have enough complexity (Note that the scenario is not designed to be a zero-sum game).

\vspace{-12pt}

\section{Conclusions and Main Implications}

\vspace{-8pt}

This work introduces the first comprehensive formalization, implementation and quantitative evaluations of \feint in Multi-Player Games. We provide automatic generation of \feint behaviors using Palindrome-directed Templates and synergistically combine \feint with follow-up actions in Dual-Behavior Model. The decision choices on the action-level are fused into strategy-level formalizations in game interactions. We provide a concrete implementation scheme to incorporate \feint into common MARL frameworks. The results show that our design of \feint can (1) greatly improve the reward gains from the game; (2) significantly improve the diversity of Multi-Player Games; and (3) only incur negligible overheads in terms of the time consumption. We conclude that our formalization of \feint is effective and practical, to make Multi-Player Games more interesting. 

The successful formation of \feint behaviors and strategies imply the potential unsafety of existing machine learning models (and, clearly, future ones also). Therefore, the wide adoption of machine learning models certainly demand the consideration of this work (and its variants), for building responsible Artificial Intelligence; and/or leveraging them for the future society in a responsible way.

\bibliographystyle{plain}
\bibliography{main-arxiv}


\newpage

\appendix

\section{Conceptual Clarifications}

Since this work spans multiple disciplines, there are a few clarifications to ensure the consistent understanding between our work and prior arts.

\subsection{Differences between actions and behaviors in our formalization}
\label{appendix:action-behavior-dif}

To provide a unified definition of \feint behavior in both continuous and discrete action space, we highlight the difference between the terms \textbf{action} and \textbf{behavior} used in our formalization. We use \textbf{action} as the minimal unit movement in a unit time step, such as a unit step movement along the X and Y axis in a 2D board game, raising arms for a certain distance in a boxing game, turning steering wheels while applying brakes for a certain degree in a racing game, etc. This definition of action coincides with the commonly used definition of action in general MARL environments, which is intuitive to understand, simulate, and build our formalization of \feint upon it. One may argue that in some game simulations, combat movements like a cross punch are simply considered as one action, but one can always divide those movements into several unified unit-time-step actions to create a unified alignment in terms of time step in games. In terms of \textbf{behavior}, we refer to it as a combination of several actions in a sequence (e.g., a cross punch in boxing games). Thus, \feint can be naturally defined as a behavior that uses a sequence of actions to deceive opponents and lead to large reward actions in the near future. We describe our observation of \feint behaviors' characteristics and introduce our formalization at the action level in Section~\ref{sec:formalization-action-level}.

\subsection{Modeling Behaviors at Action-Level in Game Animation and Simulation}
\label{appendix:feint-action}

Modeling characters' behaviors (series of actions) in games can be divided into two categories based on the main purpose: animation-driven modeling or simulation-driven modeling, though animation and simulation are inherently closely correlated. Animation-driven methods mainly focus on modeling the behaviors themselves, with goals of producing a variety of nuanced and coherent action sequences. The interactions with the environment (whether physics-based or not) are generally considered after the modeling of the behaviors and are generally simplified to showcase the behaviors themselves. \textbf{Patch-based generation} is a direct way for such methods, which directly compose behaviors by combining pre-defined action sequences~\cite{SIGGRAPH21/Two-Player-Game}. This approach is widely adopted in the industry due to its high production efficiency, supported by an extensive amount of animation libraries (e.g. Mixamo~\cite{Mixamo})~\cite{patch_based_animation1,patch_based_animation2,patch_based_animation3}. However, in recent years, \textbf{Learning-based generation} dominates the field as they can automatically produce animated behaviors to mimic the styles of learned actions from the training inputs~\cite{learning_based_animation2,learning_based_animation3}. On the other hand, simulation-driven modeling usually considers the full interactions with the environment in the first hand. These methods generally formalize the behavior modeling process using Reinforcement Learning (RL) based frameworks to fully explore the complicated space of physics-based action modeling~\cite{SIGGRAPH21/Two-Player-Game}. In our work, we use a animation-driven modeling with strong physical constraints to describe our observations of \feint behavior characteristics and use the general simulation-driven modeling in MARL schemes for learnable formalization of \feint in action and strategy levels.

\newpage

\section{Feint Behavior Generator and the Resulting Templates}

\label{appendix:feint-actions-3-stage}

Under the above three-stage decomposition of attack behaviors, there are abundant possibilities to compose \feint behaviors from the three action sequences. However, to ensure physically realistic generation, we summarize two requirements that \feint behaviors must follow: (1) \feint behaviors should follow semi-symmetrical patterns to effectively deceive opponents and return to a rest position for follow-up moves. In boxing, a human player must retract the stretched-out limbs to the relatively rest position, before stretching out to perform an actual attack action. This is because the retraction requires recharging the force to contracted muscles; and (2) transitions between adjacent actions in different behaviors are expected to be smooth, as humanoid body movements must provide continuous movements.

To satisfy the above two requirements, we propose a \feint behavior template generator called \textbf{Palindrome-directed Generation of \feint Templates}, by extracting subsets of semi-symmetrical actions from an attack behavior and synthesizing them as a \feint behavior. The general method to generate these templates are (1) by extracting subsets of unit actions from an attack behavior, a \feint behavior can be considered as a semi-finished real attack behavior. This ensures the high similarity of a generated \feint behavior with an attack behavior, thus opponents can be deceived; and (2) by synthesizing semi-symmetric action sections, the overall movements can be connected smoothly and the naturalness of humanoid actions can be guaranteed. Within our proposed template generator \textbf{Palindrome-directed Generation of \feint Templates}, there are two key adjustable parameters in practice: (1) sequence composition positions for \feint templates; and (2) sequence length for \feint templates. We provide the rationales for these two key design choices.

(1) \textbf{Sequence composition positions for \feint templates:} Determining which position to extract the subsets of action sequences needs to ensure that the extracted actions are semi-symmetrical and allow physically realistic connections. To this end, we can have three templates with different restrictions to exploit the composing patterns: (A) For template \ding{202}, if there are similar physical states, which refer to the positions of all joints and stretching angles are similar (as shown in \ding{202} of Figure~\ref{fig:three-stage}), actions before the first similar state and after the second similar state can be extracted and directly synthesized as a \feint behavior (shown in \ding{202} of Figure~\ref{fig:three-stage}); (B) For template \ding{203}, by cutting once at any time point in Sequence 1, action sequences before the selected point and the corresponding reversion can be synthesized as a \feint behavior (shown in \ding{203} of Figure~\ref{fig:three-stage}); and (C) For template \ding{204}, similar to the second situation, by cutting once at any time point in sequence 3, action sequences after the selected point and the corresponding reversion can be synthesized as a \feint behavior (as shown in \ding{204} of Figure~\ref{fig:three-stage}). With these considerations, the \feint behavior generation templates guarantee the naturalness of continuous movements via semi-symmetrical patterns.

(2) \textbf{Sequence length for \feint templates:} The choices for the length of extracted action sequences in each template can vary greatly, since multiple actions in an attack behavior can be extracted based on different time ranges. The available choices can be any time length that results in action sequences that satisfy the physical requirements discussed above (e.g. morphologically reasonable Template \ding{203} or Template \ding{204} in Figure~\ref{fig:three-stage}). Note that it is also possible to construct nested \feint behaviors, given a large number of feasible extraction positions.  We formalize this choice as a learnable parameter that needs to combine \feint behaviors with their intended follow-up actions (Section~\ref{subsec:action-Dual-Behavior-Model}), and the learning adjustment is described in Section~\ref{sec:implementation}.

\newpage

\section{Demonstration of \feint Behaviors}
\label{appendix:feint-demo}

\subsection{Demonstration of \feint Behaviors in Dual-Beahvior Models}
\begin{figure*}[h]
  \centering
\includegraphics[width=\linewidth]{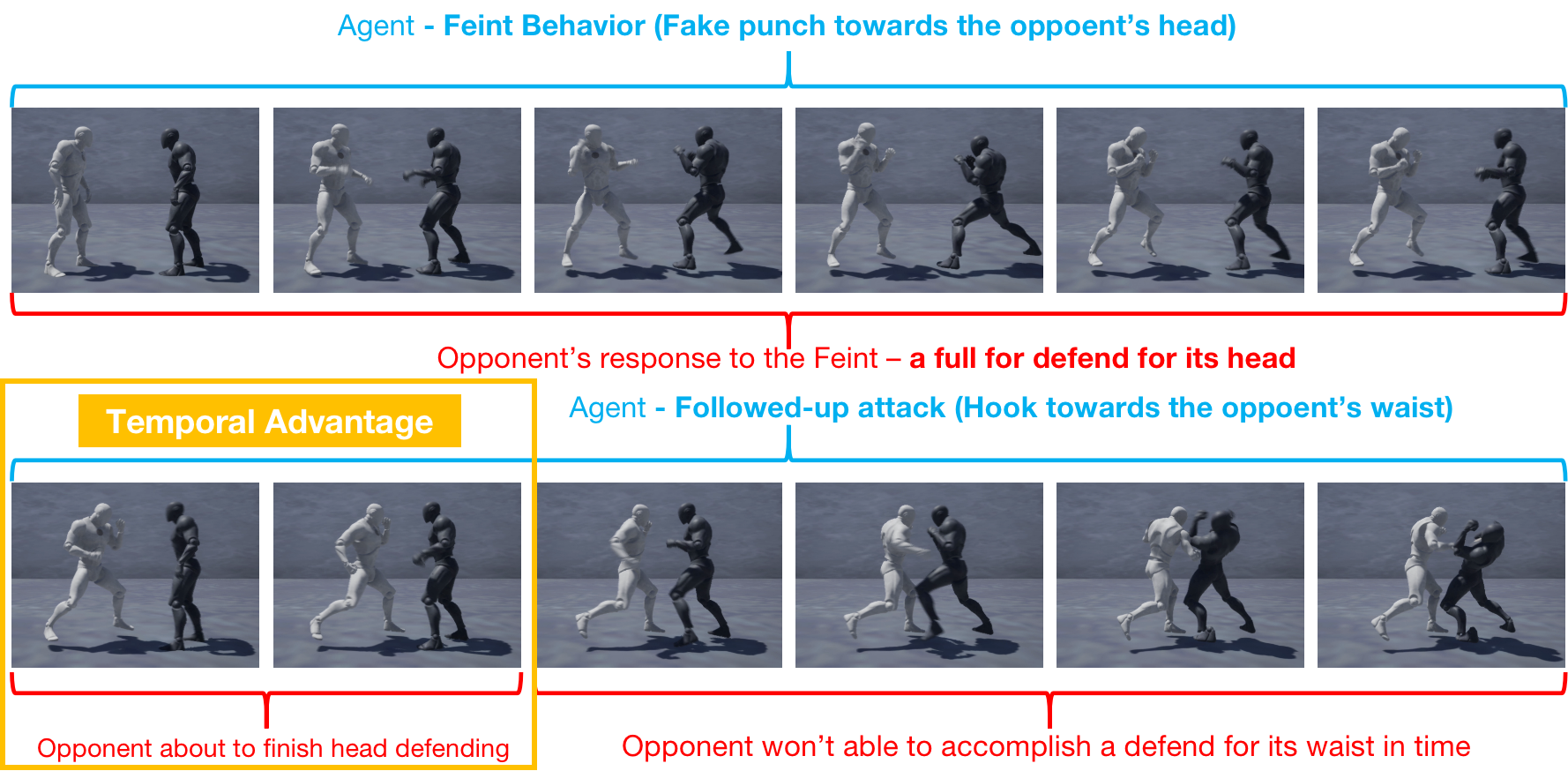}
  \caption{Dual-action Model - snapshots of the full process}
  \label{fig:dual-behavior-model-example}
\end{figure*}

To explain the generation of physically realistic \feint behavior in a Dual-Behavior Model in detail, we use humanoid models: when selecting the corresponding actions (i.e. from \feint behaviors and then an attack behavior), the starting position (jointly connected body) of the second action should be the same as the ending position of the starting action. With such a principle, the joints of a character's body can perform natural movements during the transition between these two behaviors. Figure~\ref{fig:dual-behavior-model-example} demonstrates a physically realistic combination of a \feint behavior and a follow-up attack behavior. When checking the end of NPC A's \feint behavior and the beginning of the Agent's (left white agent) real attack, both the upper and lower body parts of NPC A perform the same postures (the left arm raised and the right arm charged, performing a punch for the upper body, and the left foot forward for lower body).

Figure~\ref{fig:dual-behavior-model-example} provides a detailed example of a successful \feint behavior in a Dual-Behavior Model. We refer to the Agent as the white player on the left and its Opponent as the black player on the right, and describe the \feint behavior from the Agent perspective. The agent first performs a \feint behavior which is fake punch towards its opponent's head, which leads the opponent to defend towards its head. However, the agent connects such \feint behavior with a follow-up hook towards the opponent's waist. Due to the temporal advantage gained by the quick \feint behavior and the spatial advantage gained by deceiving the opponents to defend to wrong directions, the opponent would be knocked down by the follow-up behavior of the agent. Thus, a successful \feint behavior is performed in this Dual-Behavior Model.

\subsection{Demonstration of Successful and Unsuccessful \feint Behaviors}

\begin{figure}[htbp]
    \centering
    \includegraphics[width=\linewidth]{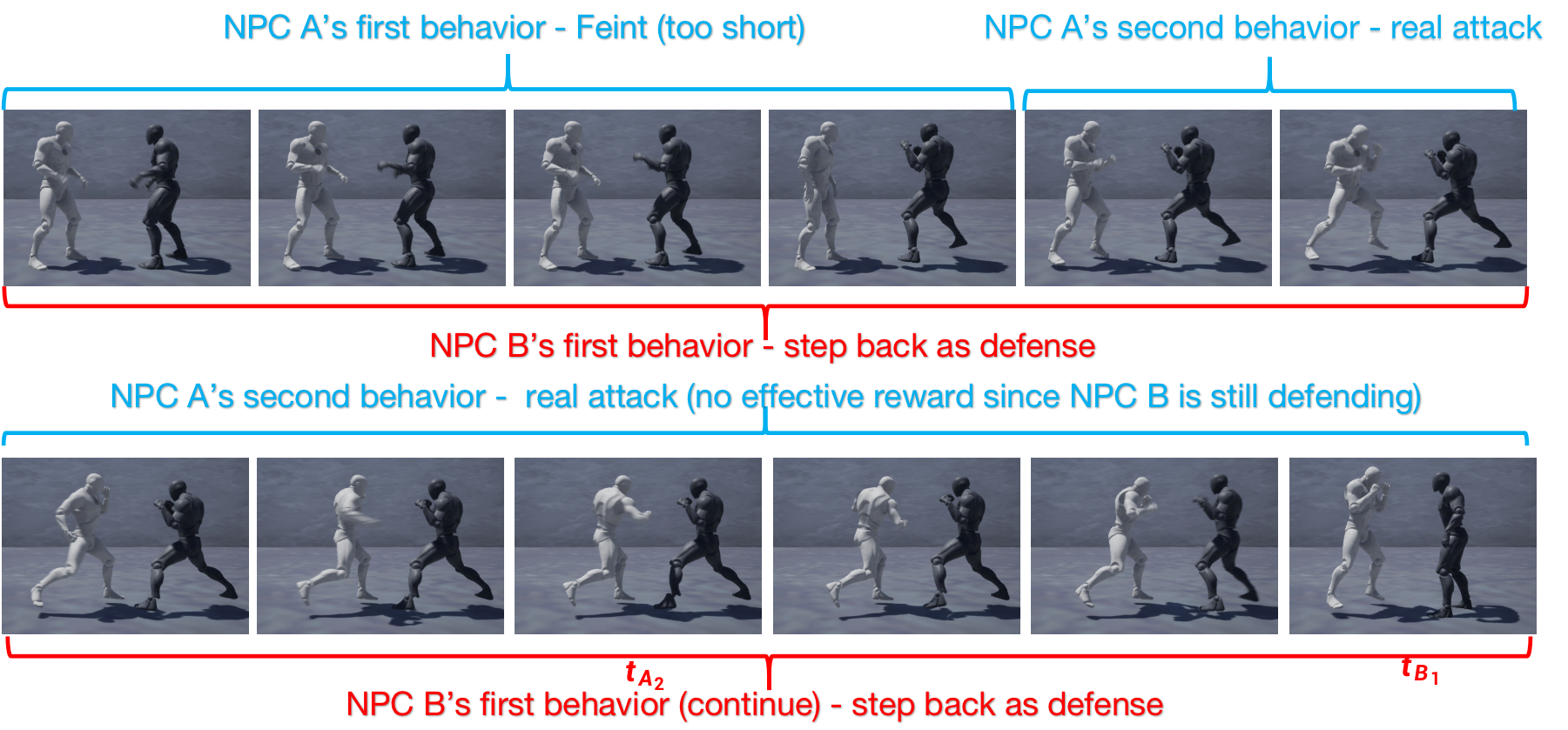}
    \caption{Demonstration of unsuccessful \feint behavior when its too short}
    \label{fig:feint-too-short}
\end{figure}

\begin{figure}[htbp]
    \centering
    \includegraphics[width=\linewidth]{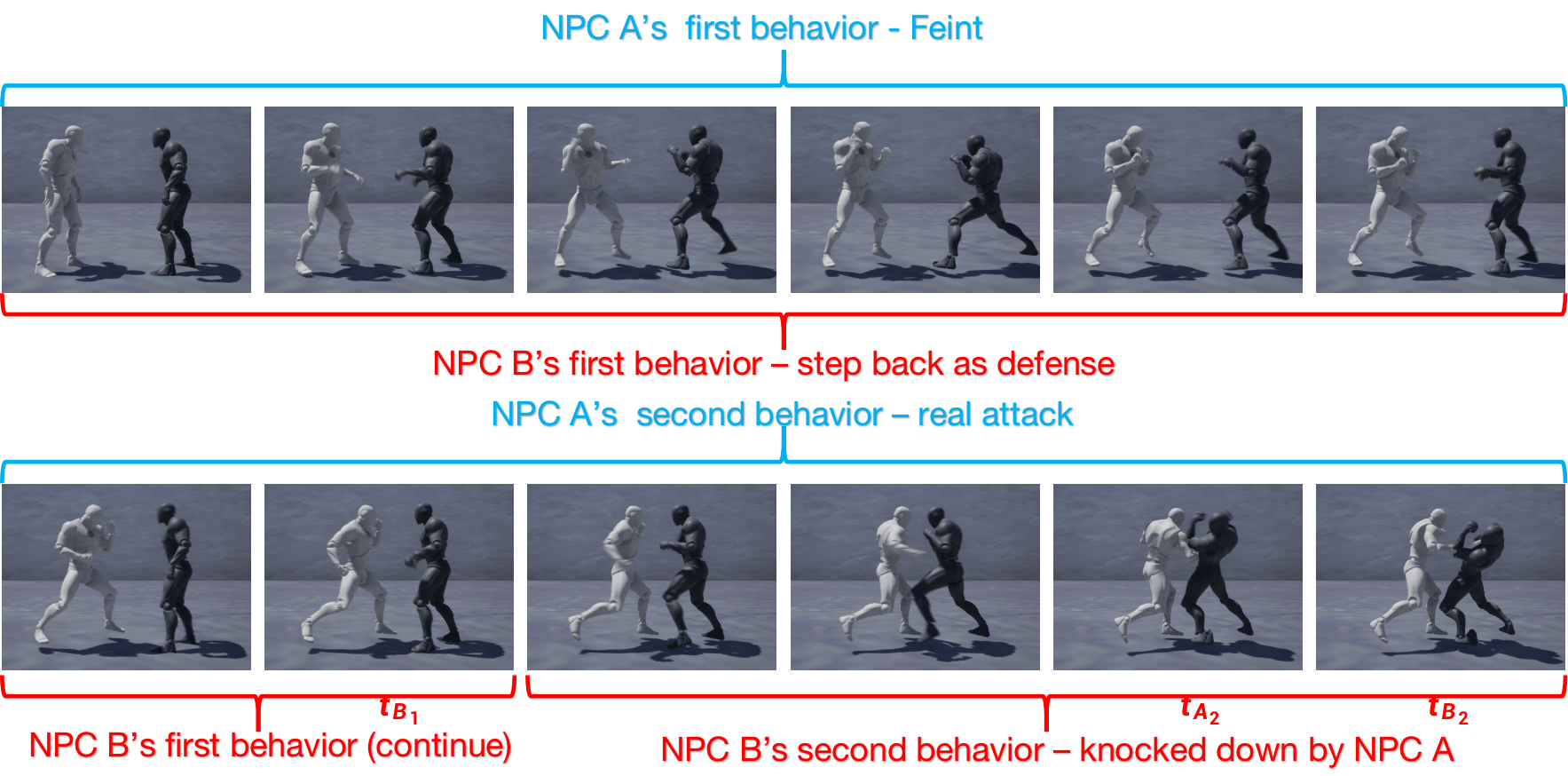}
    \caption{Demonstration of successful \feint behavior with proper length}
    \label{fig:feint-proper}
\end{figure}

\begin{figure}[htbp]
    \centering
    \includegraphics[width=\linewidth]{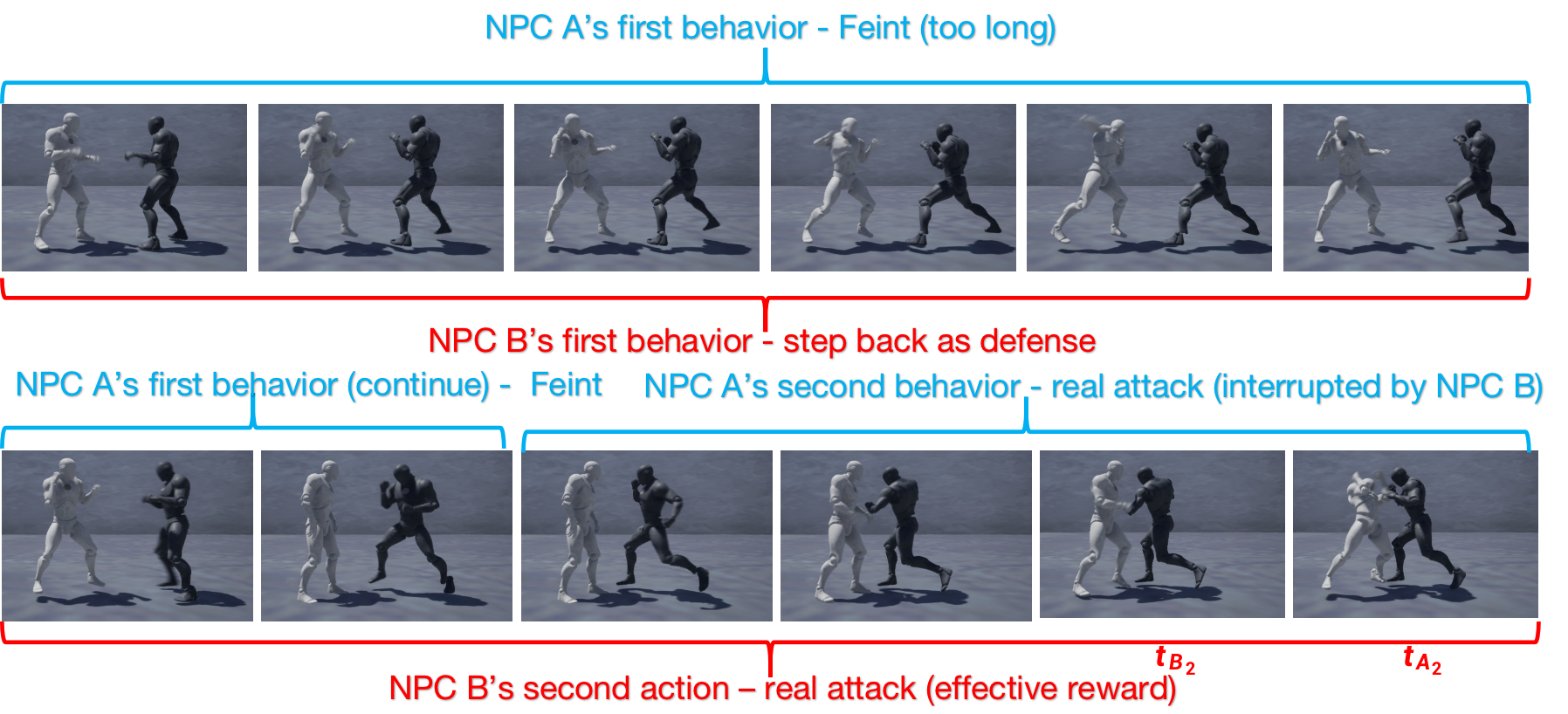}
    \caption{Demonstration of unsuccessful \feint behavior when its too long}
    \label{fig:feint-too-long}
\end{figure}

To enable a successful \feint behavior in a Dual-Behavior Model, the temporal and spatial advantages should be properly formalized. The advantages of combining \feint behaviors with follow-up high-reward actions stem from an appropriate time difference, incurred by \feint behaviors to mislead the opponents' actions. If the length of a \feint behavior is too short, the following attack actions might not gain much advantage compared to actions combinations without \feint behaviors; and if the length of a \feint action is too long, the process to perform a \feint behaviors can leave sufficient time for the opponent to react and even attack back. We provide examples for these scenarios in Figure~\ref{fig:feint-too-short}, Figure~\ref{fig:feint-proper}, and Figure~\ref{fig:feint-too-long}. We refer to the left white player as NPC A and describe the \feint from its perspective, and the right black agent NPC B is considered as its opponent.

We use the timeline of the Dual-Behavior Model in Figure~\ref{fig:dual-behavior-model-abstract} to analyze and evaluate the three \feint behaviors. We use three key time points that are highlighted in Figure~\ref{fig:feint-too-short}, Figure~\ref{fig:feint-proper}, and Figure~\ref{fig:feint-too-long} to explain the action sequences, in which \begin{math}t_{B_1}\end{math} indicates the end of defense behavior while \begin{math}t_{A_2}\end{math} indicates the estimated start of reward in the second action sequence for NPC A and \begin{math}t_{B_2}\end{math} indicates the estimated start of reward in second action for NPC B. The three consequences mainly differ in these three key time points.

1) \textbf{Very short \feint behaviors \begin{math}t_{A_2} < t_{B_1}\end{math}:} The action sequence of simulation is shown in Figure~\ref{fig:feint-too-short}, in which the \feint behaviorsduration is extremely short and the estimated start of reward in second action for NPC A (\begin{math}t_{A_2}\end{math}) happens when NPC B is still in the first defense action (thus \begin{math}t_{A_2} < t_{B_1}\end{math}). As the sequence shows, the second real action of NPC A would not benefit much since NPC B is still in defense.

2) \textbf{Proper length \feint behaviors \begin{math}t_{B_1} < t_{A_2} < t_{B_2}\end{math}:} The action sequence of simulation is shown in Figure~\ref{fig:feint-proper}, in which the \feint behaviors have a moderate duration. The key difference of this duration is that the estimated start of reward in the second behavior for NPC A happens after the end of the defense behavior of NPC B and before the estimated start of reward in the second behavior for NPC B, thus showing the temporal advantages introduced in Section~\ref{subsec:action-Dual-Behavior-Model}. With such temporal advantages, NPC A gains preemptive advantage over NPC B, inflicting rewards from NPC B (at time \begin{math}t_{A2}\end{math} in Figure~\ref{fig:feint-proper}) before NPC B's reward inflicting of second behavior starting (at time \begin{math}t_{B2}\end{math} in Figure~\ref{fig:feint-proper}). When NPC A hits NPC B at \begin{math}t_{A2}\end{math}, the ongoing action of NPC B will be interrupted and NPC B would be knocked down.

3) \textbf{Very long \feint behaviors \begin{math}t_{A_2} > t_{B_2}\end{math}:} The action sequence of simulation is shown in Figure~\ref{fig:feint-too-long}, in which the \feint actions duration is too long and the estimated start of reward in second behavior for NPC A (\begin{math}t_{A_2}\end{math}) happens after the estimated start of damage in second action for NPC B (\begin{math}t_{B_2}\end{math}). This condition has the opposite consequence of a moderate length \feint behaviors, in which NPC B can inflict rewards on NPC A before NPC A's reward inflicting of the second behavior starts. When NPC B hits NPC A at \begin{math}t_{A2}\end{math}, the ongoing action of NPC A will be interrupted and NPC A would be knocked down.

Thus, the choice of the time duration for Feint actions highly depends on the action combinations and the estimation of opponents' actions, proving our observation in Section~\ref{sec:formalization-action-level}. Thus the learning to formalize such a choice in the strategy learning scheme (Section~\ref{sec:formalization-strategy-level}) is important to construct effective \feint behaviors with corresponding Dual-Behavior Models.

\newpage

\section{Testbed Implementations}
\label{appendix:testbed}

Our main testbed game environment is a multi-player boxing game, which is based on OpenAI's open-source environment Multi-Agent Particle Environment \cite{MA-Partical-Env}, but with heavy additional implementation to create a physically realistic scenario.This game resembles intense free fight scenarios in ancient Roman free fight scenarios~\cite{Roman-Fight}, where interactions are intense and \feint is expected to be effective. We incorporate common boxing behaviors (action sequences) in boxing games. following the methodology in some animation and simulation works~\cite{SIGGRAPH10/Animation-Two-Player-Game,SIGGRAPH21/Control-Strategy-Two-Player-Game}. This handcrafted scenario contains complex physics-based interaction systems and fine-grained time steps to enable learning and generating \feint behaviors. A detailed description of the reward gaining system, environment parameters, and agent settings is presented in Appendix~\ref{appendix:scenario-detail}. We also modify and extend a strategic real-world game, AlphaStar~\cite{GECCP19/AlphaStar}, which is widely used as the experimental testbed in recent studies of Reinforcement Learning studies~\cite{Journal/AlphaStar,NIPS21/BD-RD}. We make extra efforts to emulate a six-player game, where players are free to have convoluted interactions with each other. And we implement \feint as dynamically generated policies, based on the 888 regular gaming policies.

\subsection{Details of Boxing Game Scenario}
\label{appendix:scenario-detail}

Our testbed game scenario is emulates a complex boxing game by modeling all the detailed combat behaviors except building the graphical rendering process. The reason we neglect the rendering process is that our main goal is to evaluate the effectiveness of formalization of \feint behaviors in multi-player games, and the building a real-time graphical rendering with such complex humanoid interactions would be a graphics paper itself. We fully emulate all the behavior details in our game simulation, thus our constructed game simulation is detailed enough to evaluate our formalization of \feint behaviors. We provide a detailed description of the game scenario here.

\begin{figure}[h]
    \centering
    \includegraphics[width=\linewidth]{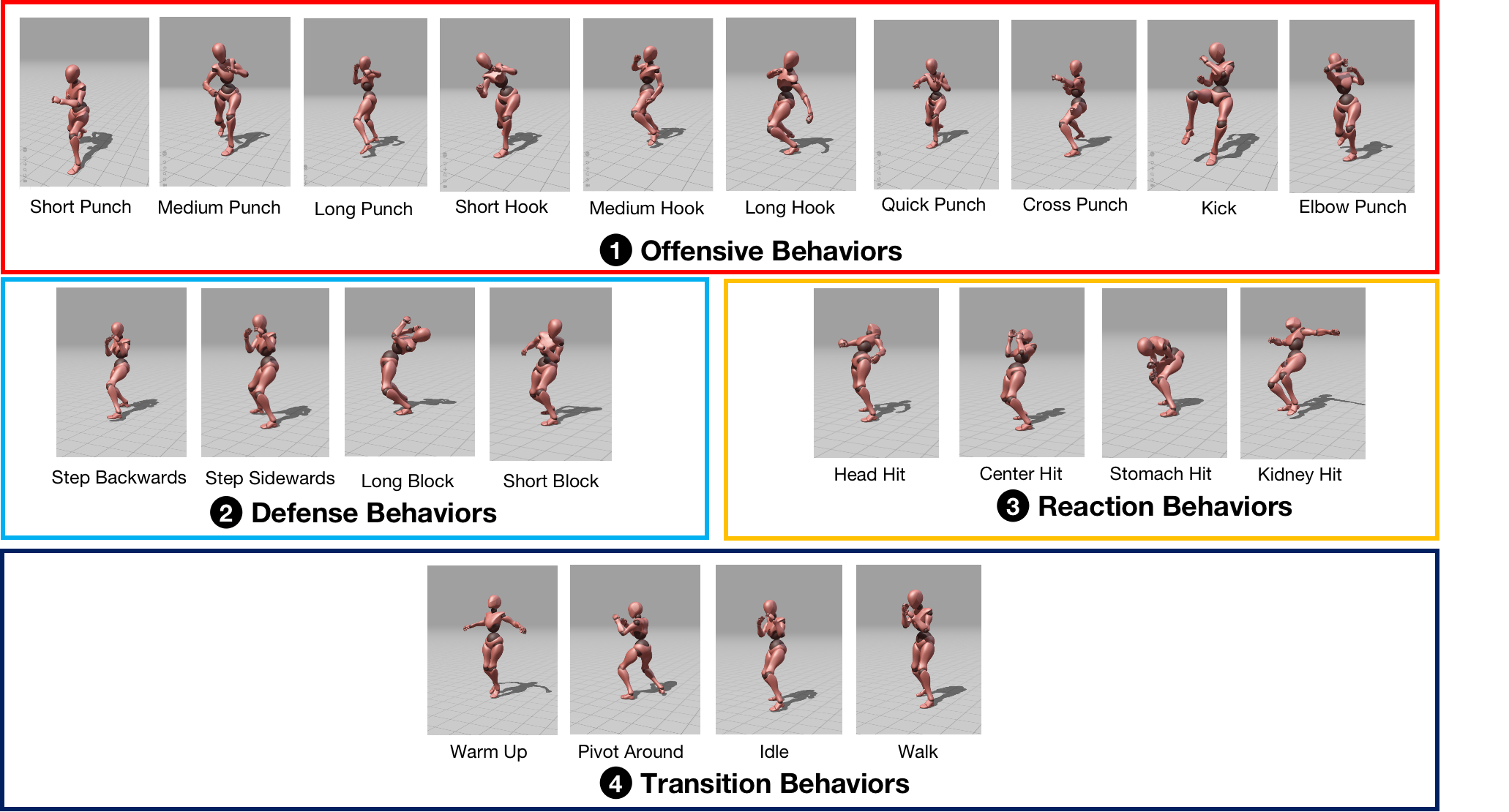}
    \caption{The full set of 22 behavior (action sequences) of a boxing game from Mixamo.}
    \label{fig:mixamo-set}
\end{figure}

We follow a similar boxing game scenario construction approach as~\cite{SIGGRAPH10/Animation-Two-Player-Game, SIGGRAPH21/Two-Player-Game}, and model the full set of Mixamo~\cite{Mixamo} 22 behaviors (action sequences) which contain over 250 available full body actions (illustrated in Figure~\ref{fig:mixamo-set}). We extensively construct a gaming environment based on Multi-Agent Particle System~\cite{MA-Partical-Env} to incorporate these behaviors, which then can be seamlessly integrated with common MARL models.

\begin{figure}[h]
    \centering
    \includegraphics[width=\linewidth]{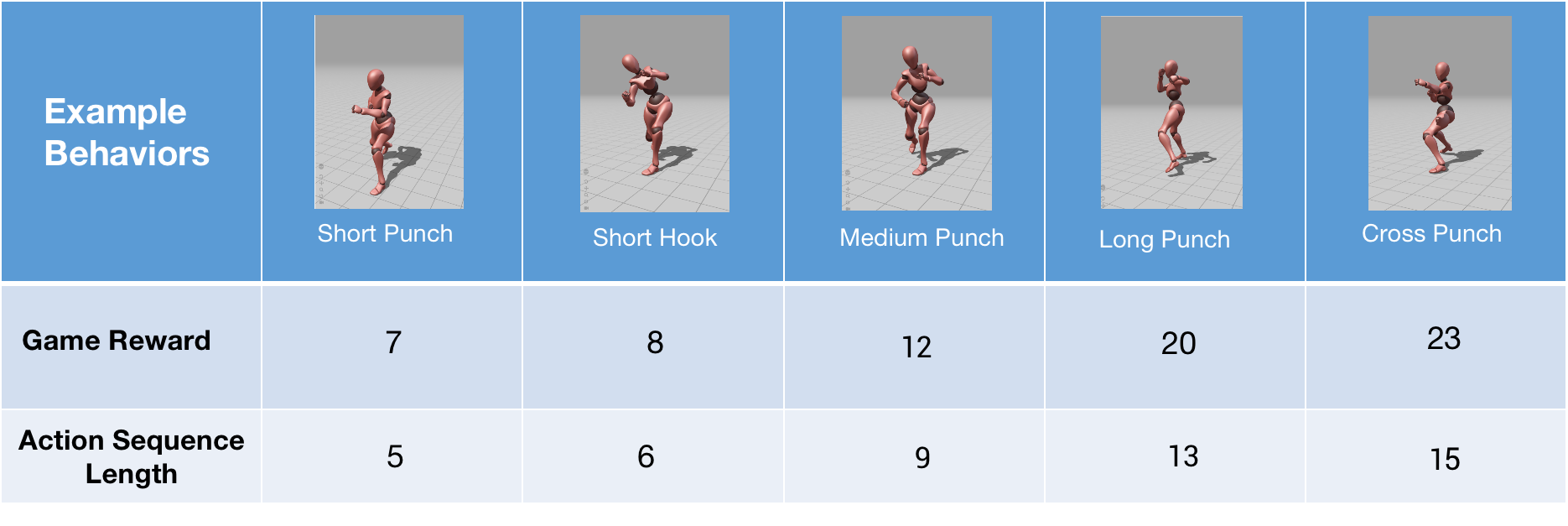}
    \caption{Demonstration of the game rewards and action sequence lengths of 5 Mixamo behaviors.}
    \label{fig:mixamo-length-reward}
\end{figure}

The players can move around in a 2D plane. We use a vector to model the physical state of players, which stores and tracks the body movements of a player. This vector tracks the positions of body parts: left and right limbs, the left and right legs, and the center body, which is used to select available combat behaviors (the transitions of body movements must be smooth as mentioned in Section~\ref{subsec:action-Feint-templates} and Section~\ref{subsec:action-Dual-Behavior-Model}). With this setting, \feint behaviors can be naturally generated and incorporated into suitable Dual-Behavior Models. We follow the exact Mixamo dataset to model the length of the behaviors (the length of action sequences) and rewards the behaviors (e.g., a successful long punch would gain more rewards than a short punch.) Specifically, we measure the number of frames contained in all behaviors and normalize them to define unit time steps for action space and thus get the action sequence lengths for all behaviors. An example of game rewards and action sequence length of 5 behaviors are provided in Figure~\ref{fig:mixamo-length-reward}. 

\subsection{Experimental Procedure}
\label{appendix:experiment-procedure}
We choose 4 commonly used MARL models: MADDPG~\cite{NIPS17/MADDPG}, MASAC~\cite{ICML18/SAC, ICML19/MAAC}, MATD3~\cite{NIPS19/MATD3}, and MAD3PG~\cite{ICLR18/D4PG,MDPI21/MAD3PG} and incorporate them into testbed scenarios. Our implementation is based on~\cite{20/implementaion}, which provides a unified MARL frameworks for the above models. We aim to test whether \feint behaviors can be uniformly and effectively learned using all these commonly used MARL models and how can \feint affect the game rewards for agents. Note that our purpose is to verify the effectiveness of our formalization of \feint behaviors and not to compare or modify the MARL models themselves. We create two test scenarios, the first one with two players (one player per team) and the second one with six players (3 players per team). For all of these scenarios, we first train the agents without \feint as baselines using the 4 models. Then for the two-player scenario, we incorporate \feint on one player (shown as the Good player in Figure~\ref{fig:1_vs_1}). For the six-player scenario, we select 1 agent in the Good team (labeled as Good 3 in Figure~\ref{fig:3_vs_3}), to incorporate our formalization of \feint, and keep all other 3 agents regular. The reason for this design in the six-player scenario is that we want to not only test how \feint behaviors can affect the reward gain against direct opponents, but also see whether \feint can bring advantages for a player among its teammates. All the players are rewarded independently and the notion of the "Good" and "Adv" team does not mean that teammates have a shared reward (i.e., not explicit constraints that force them to cooperate). Note that all players have identical capabilities and are rewarded using the same mechanisms, thus \feint can be incorporated on any player. Our labeling choice here is to provide to a consistent way to track and analysis the game rewards. All experiments for the two-player scenario are trained for 75,000 game iterations and all experiments for the six-player scenario are trained for 150,000 game iterations.

\newpage

\section{Implementation Details}
\label{appendix:implementation}

We provide the pseudo-code for computing \feint behavior templates and composing Dual-Behavior Models for our proof-of-concept implementation discussed in Section~\ref{sec:implementation}.

\begin{algorithm}
\caption{Precompute Available Feint Behavior Templates}
\label{algo::feint-templates}
\begin{algorithmic}[1]
   \State \textbf{Input:} Set of behaviors $B = \{Behavior_i | i \in [1, n]\}$, where each $Behavior_i = \{A_{i_1}, A_{i_2}, \dots, A_{i_k}\}$ is a sequence of actions.
   \State \textbf{Output:} Set of physically available Feint behavior templates $T$.
   \State Compose a set of behavior pairs: $B_{pair} = \{(Behavior_i, Behavior_j) | i, j \in [1, n]\}$
   \State Initialize empty set of templates: $T = \emptyset$
   \For {each $(Behavior_i, Behavior_j) \in B_{pair}$}
        \State Compute the common action set: $A_{common} = \{a_k | a_k \in Behavior_i \text{ and } a_k \in Behavior_j\}$
        \If {$|A_{common}| \neq 0$}
            \For {each $a_k \in A_{common}$}
                \State Available actions from $Behavior_i$: $Avail_i = \{a_m | a_m \in Behavior_i, m < k\}$
                \State Available actions from $Behavior_j$: $Avail_j = \{a_n | a_n \in Behavior_j, n > k\}$
                \State Update $T$: $T \gets T \cup \{(a_k, Avail_i, Avail_j)\}$
            \EndFor
        \EndIf
   \EndFor
   \State \textbf{return} $T$
\end{algorithmic}
\end{algorithm}

\begin{algorithm}
\caption{Backward Search to Compose Dual Behavior Models}
\label{algo:dual-behavior-models}
\begin{algorithmic}[1]
    \State \textbf{Input:} Last action $a_t$, beginning of the high-rewards behavior $a_{target}$, precomputed Feint templates $T$
    \State \textbf{Output:} Set of available Dual Behavior Models $D$
    \State Initialize empty set of Dual Behavior Models: $D = \emptyset$
    \For {each $(a_k, \text{Avail}_i = \{a_p | p \in [0, k-1]\}, \text{Avail}_j = \{a_q | q \in [k+1, max]\})$ in $T$}
        \\\Comment{Here, max is the maximum index of action in $\text{Avail}_j$}
        \If {($a_t \in \text{Avail}_i$) and ($a_{target} \in \text{Avail}_j$)}
            \State Select actions from $a_t$ to $a_k$ in $\text{Avail}_i$: $Select_i = \{a_m | m \in [t, k-1]\}$
            \State Select actions from $a_k$ to $a_{target}$ in $\text{Avail}_j$: $Select_j = \{a_n | n \in [k+1, max]\}$
            \State Update $D$: $D \gets D \cup \{(Select_i, a_k, Select_j)\}$
        \EndIf
    \EndFor
    \State \textbf{return} $D$
\end{algorithmic}
\end{algorithm}

\newpage

\section{Additional Experimental Results}
\label{appendix:additional-results}

We report the effects of \feint on \ding{202} diversity gain of policy space; and \ding{203} overhead of computation load. We examine the effects of \feint{} actions on how \feint{} can improve the diversity of gaming policies (Section~\ref{subsubsec:strategy-collective}). We also perform overhead analysis, incurred by fusing \feint formalization in strategy learning.

\subsection{Diversity Gain}
\label{subsubsec:experiment-results-diversity}
To examine the impacts on the policy diversity in AlphaStar games, we perform a comparative study between MARL training with and without \feint. Specifically, We use Exploitability and Population Efficacy (PE) to measure the diversity gain in the policy space. Exploitability \cite{NIPS17/Exploitability} measures the distance of a joint policy chosen by the multiple agents to the Nash equilibrium, indicating the gains of players compared to their best response. The mathematical expression of Exploitability is expressed as:
\begin{equation}
    Expl(\pi) = \sum_{i=1}^N(max_{\substack{\pi_i'}}Rew_i(\pi_i',\pi_{-i})-Rew_i(\pi_i',\pi_{-i}))
\end{equation}
where $\pi_i$ stands for the policy of agent $i$ and $\pi_{-i}$ stands for the joint policy of other agents. $Rew_i$ denotes our formalized Reward Calculation Model (Section~\ref{subsubsec:strategy-collective}). Thus, small Exploitability values show that the joint policy is close to Nash Equilibrium, showing higher diversity. In addition, we also use Population Efficacy (PE) \cite{NIPS21/BD-RD} to measure the diversity of the whole policy space. PE is a generalized opponent-free concept of
Exploitability by looking for the optimal aggregation in the worst cases, which is expressed as:
\begin{equation}
    PE(\{\pi_i^k\}_{k=1}^N) = min_{\substack{\pi_{-i}}} max_{\substack{1^\top\alpha=1\ a_i>=0}} \sum_{k=1}^N \alpha_k Rew_i(\pi_i^k, \pi_{-i})
\end{equation}
where $\pi_i$ stands for the policy of agent $i$ and $\pi_{-i}$ stands for the joint policy of other agents. $\alpha$ denotes an optimal aggregation where agents owning the population optimizes towards. $Rew_i$ denotes our formalized Reward Calculation Model (Section~\ref{subsubsec:strategy-collective}) and opponents can search over the entire policy space. PE gives a more generalized measurement of diversity gain from the whole policy space.

\begin{figure}[h!]
    \centering
    \includegraphics[width=0.8\linewidth]{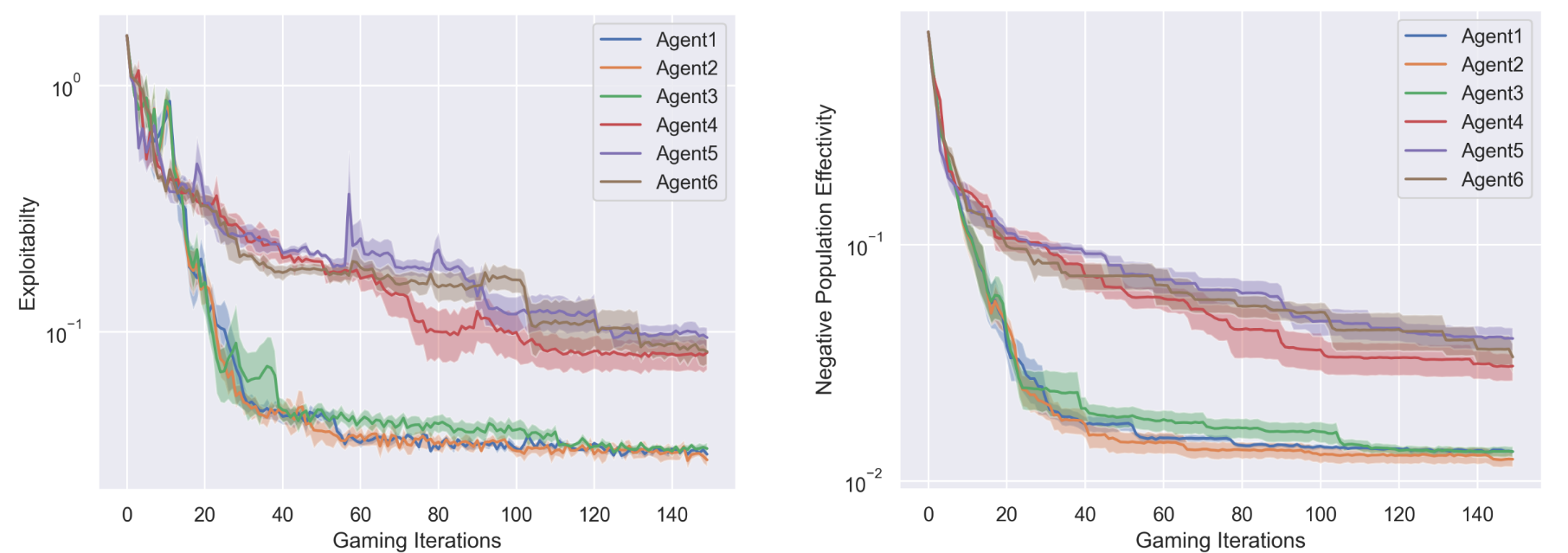}
    \caption{Diversity gain for agents, in terms of the exploitablity and the negative population efficacy.}
    \label{fig:experiment-diversity}
\end{figure}

Figure \ref{fig:experiment-diversity} shows the experimental results for evaluating diversity gains. From the figure, we obtain two observations. First, agents that can dynamically perform \feint actions (Agent 1, 2, and 3) achieve lower Exploitability (around $4.9 \times 10^{-2}$) compared to agents who perform regular actions (around $9.7 \times 10^{-2}$) and have higher PE (lower negative PE - around $5.3 \times 10^{-2}$) than those who only perform regular actions (around $1.2 \times 10^{-2}$). This result shows that our formalized \feint can effectively increase the diversity and effectiveness of policy space. Second, agents with \feint have slightly higher variations in both metrics. This is because \feint naturally incurs more randomness (e.g. succeed or not) in games, resulting in higher variations in metrics.

\subsection{Overhead Analysis}
\label{subsubsec:experiment-results-overhead}

\begin{figure}[h]
    \parbox{.03\linewidth}{\rotatebox{90}{\centering 1 VS 1}}\hfill\hfill
    \parbox{.24\linewidth}{\includegraphics[width=\linewidth]{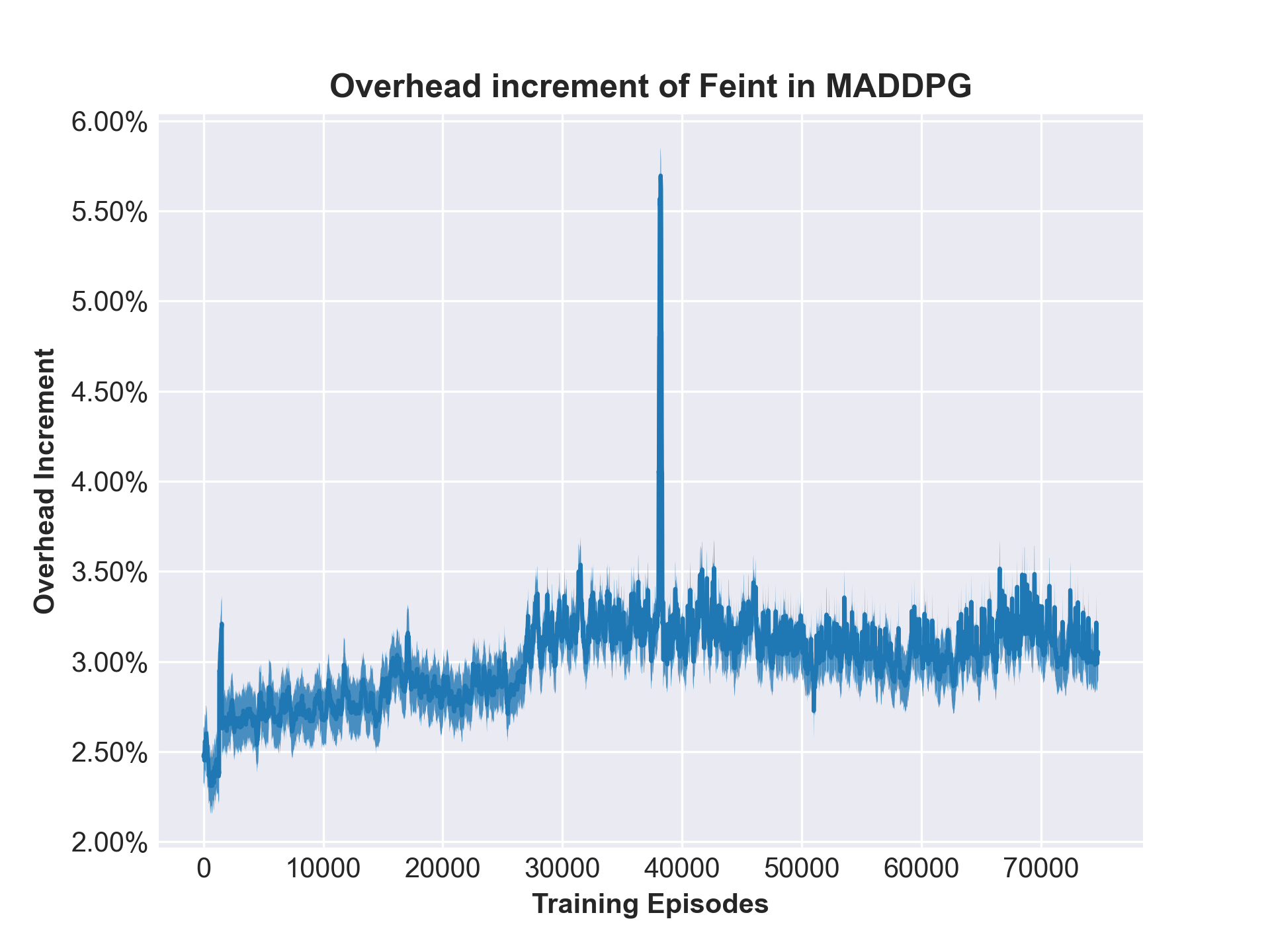}}\hfill
    \parbox{.24\linewidth}{\includegraphics[width=\linewidth]{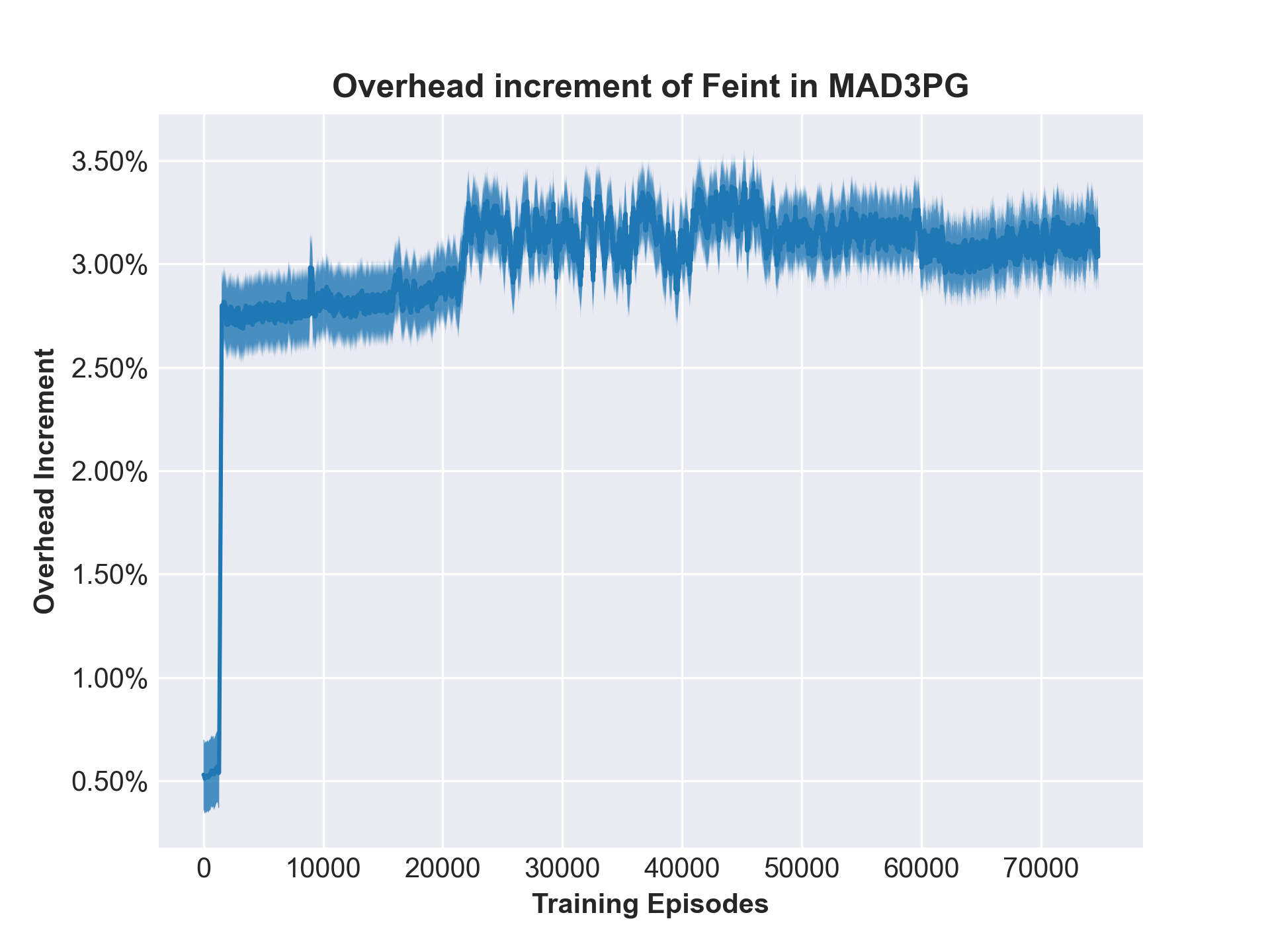}}\hfill
    \parbox{.24\linewidth}{\includegraphics[width=\linewidth]{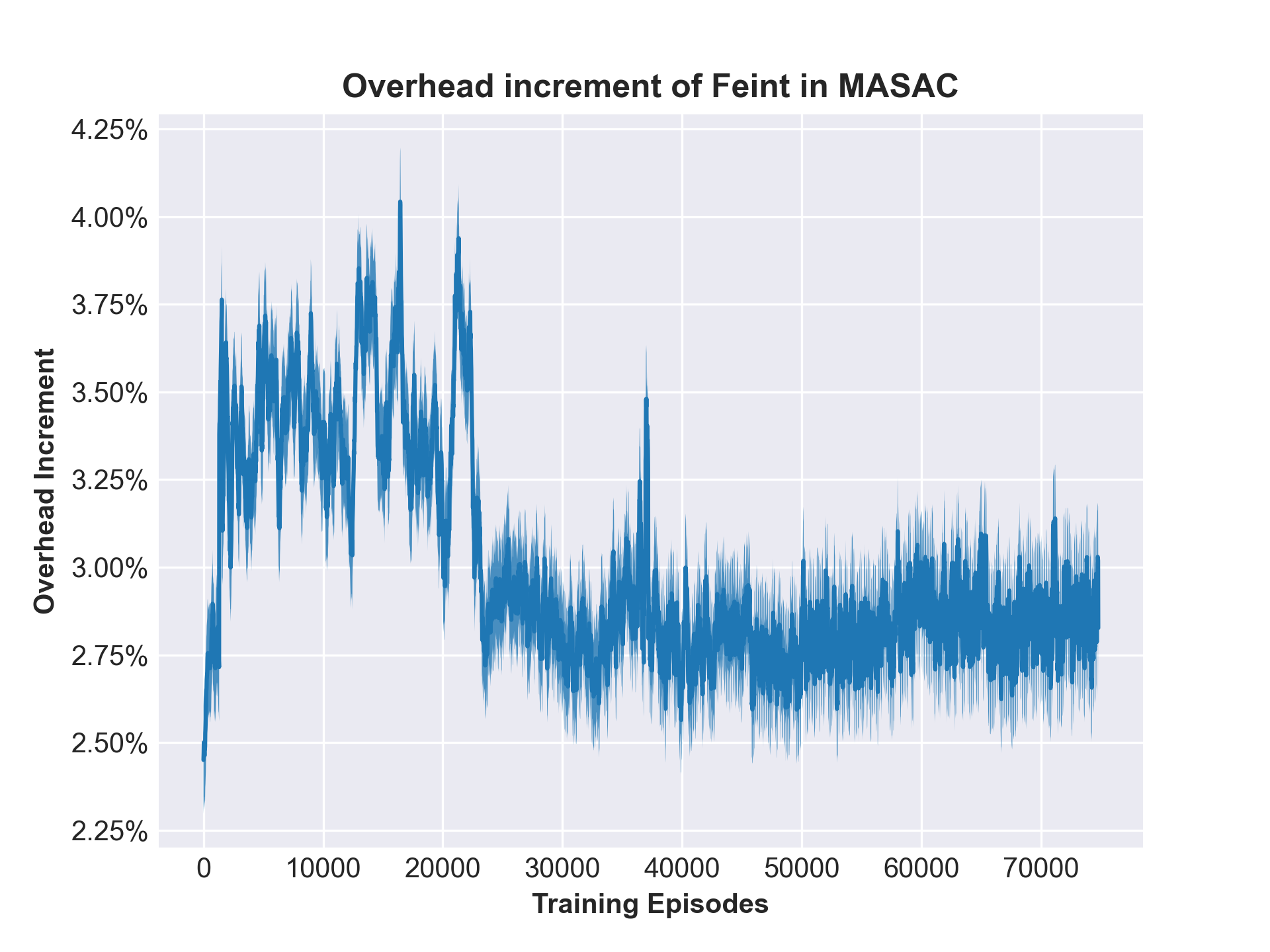}}\hfill
    \parbox{.24\linewidth}{\includegraphics[width=\linewidth]{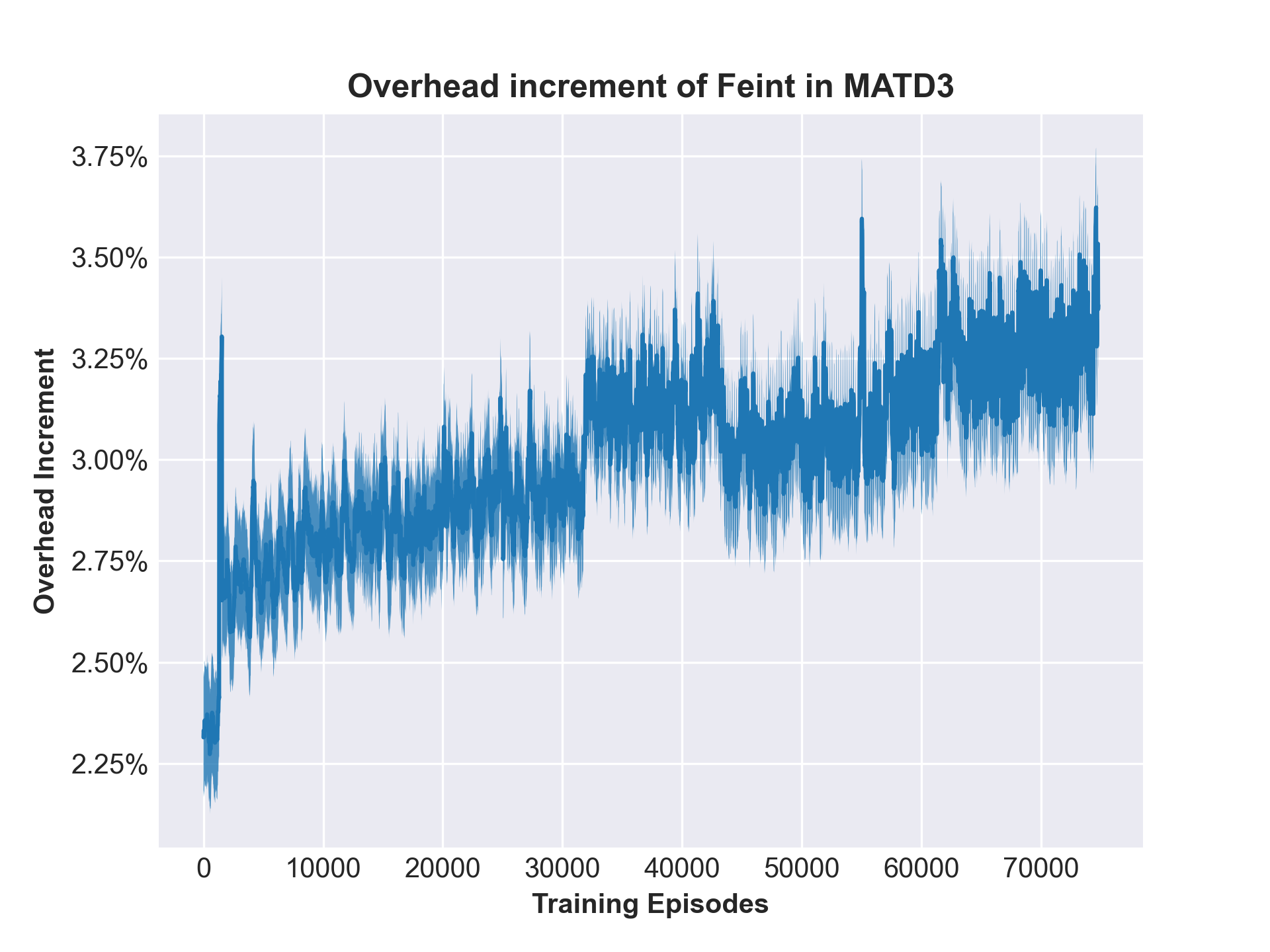}}\par
    \parbox{.03\linewidth}{\rotatebox{90}{\centering 3 VS 3}}\hfill\hfill
    \parbox{.24\linewidth}{\includegraphics[width=\linewidth]{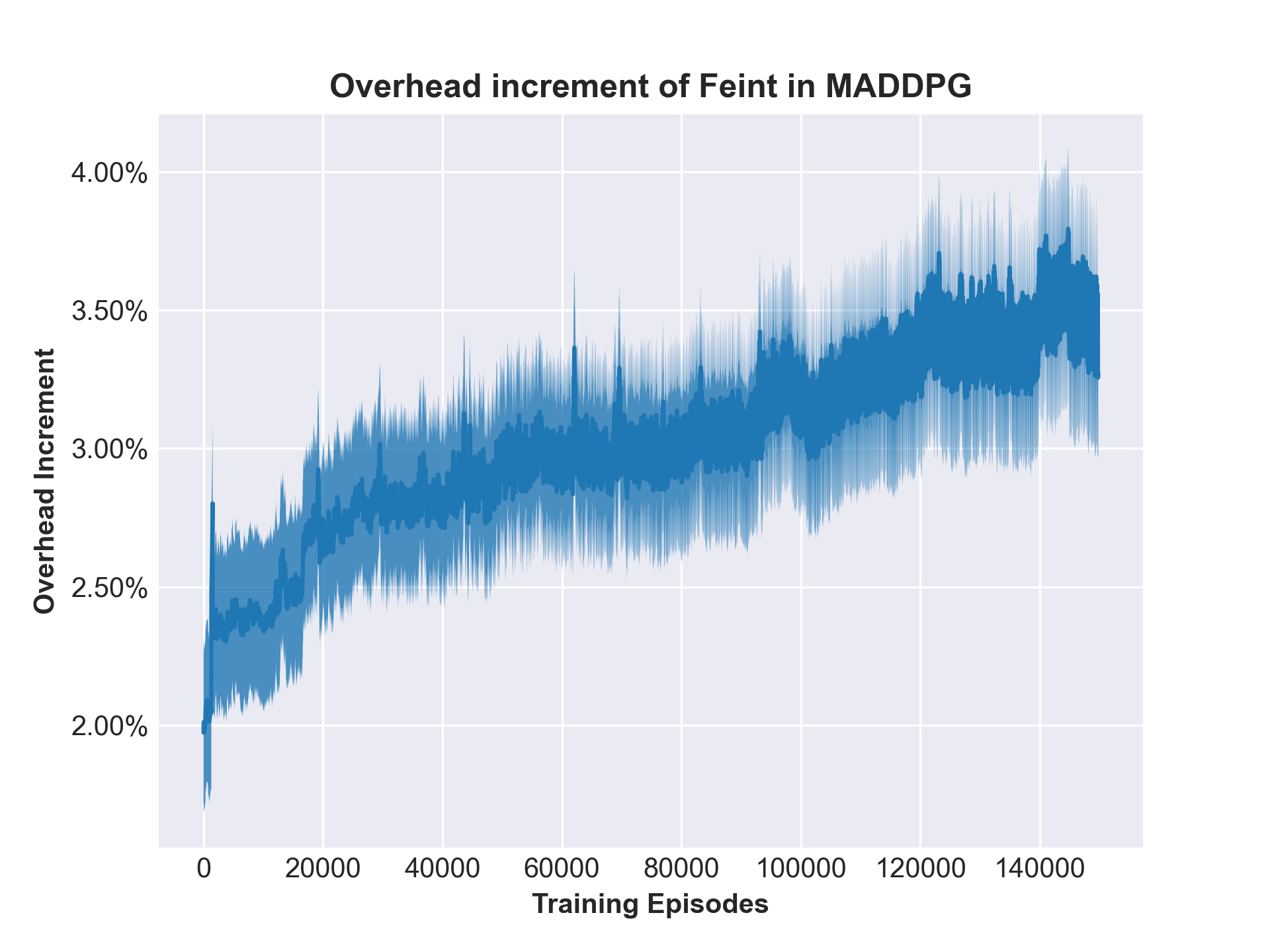}}\hfill
    \parbox{.24\linewidth}{\includegraphics[width=\linewidth]{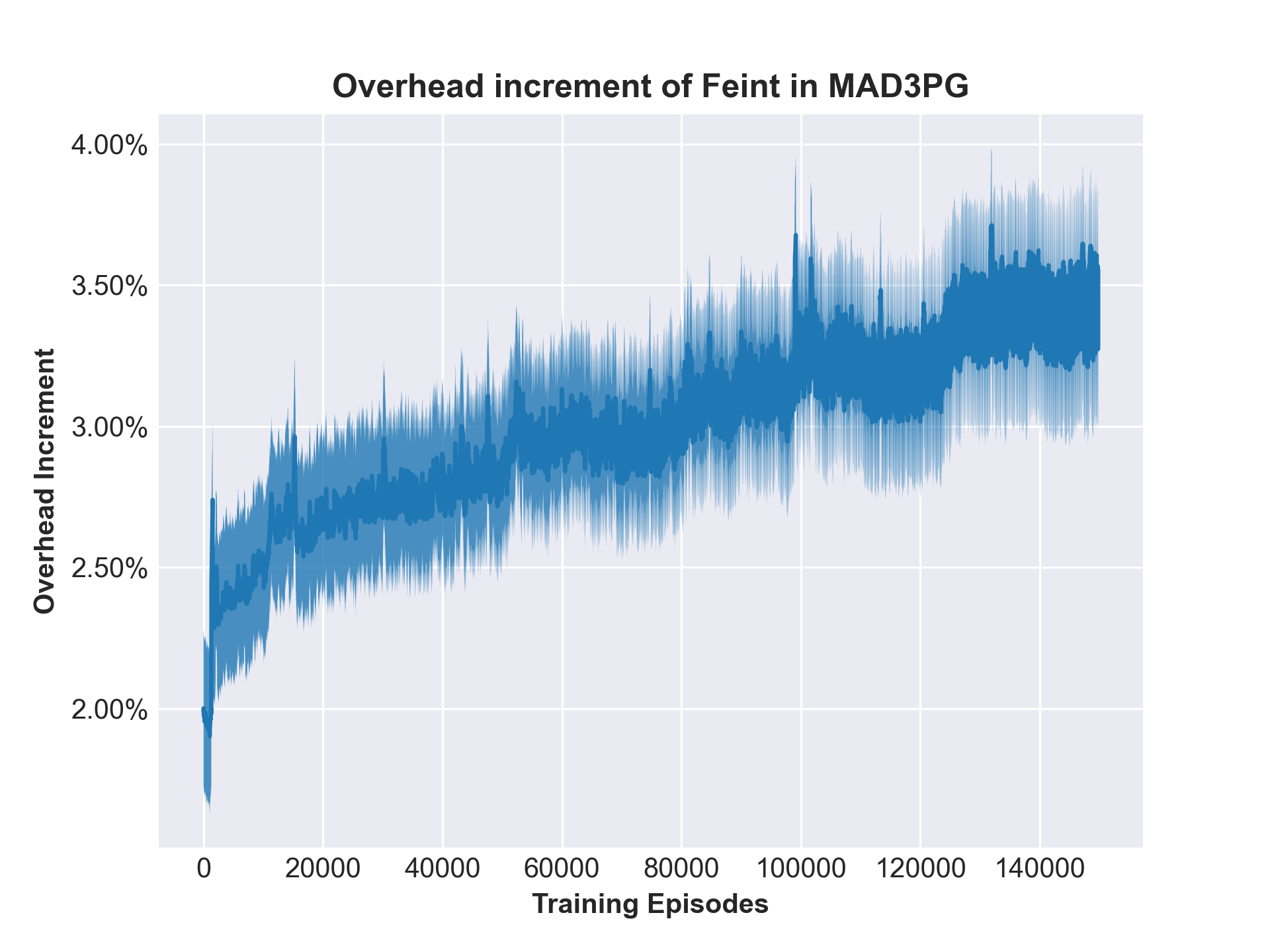}}\hfill
    \parbox{.24\linewidth}{\includegraphics[width=\linewidth]{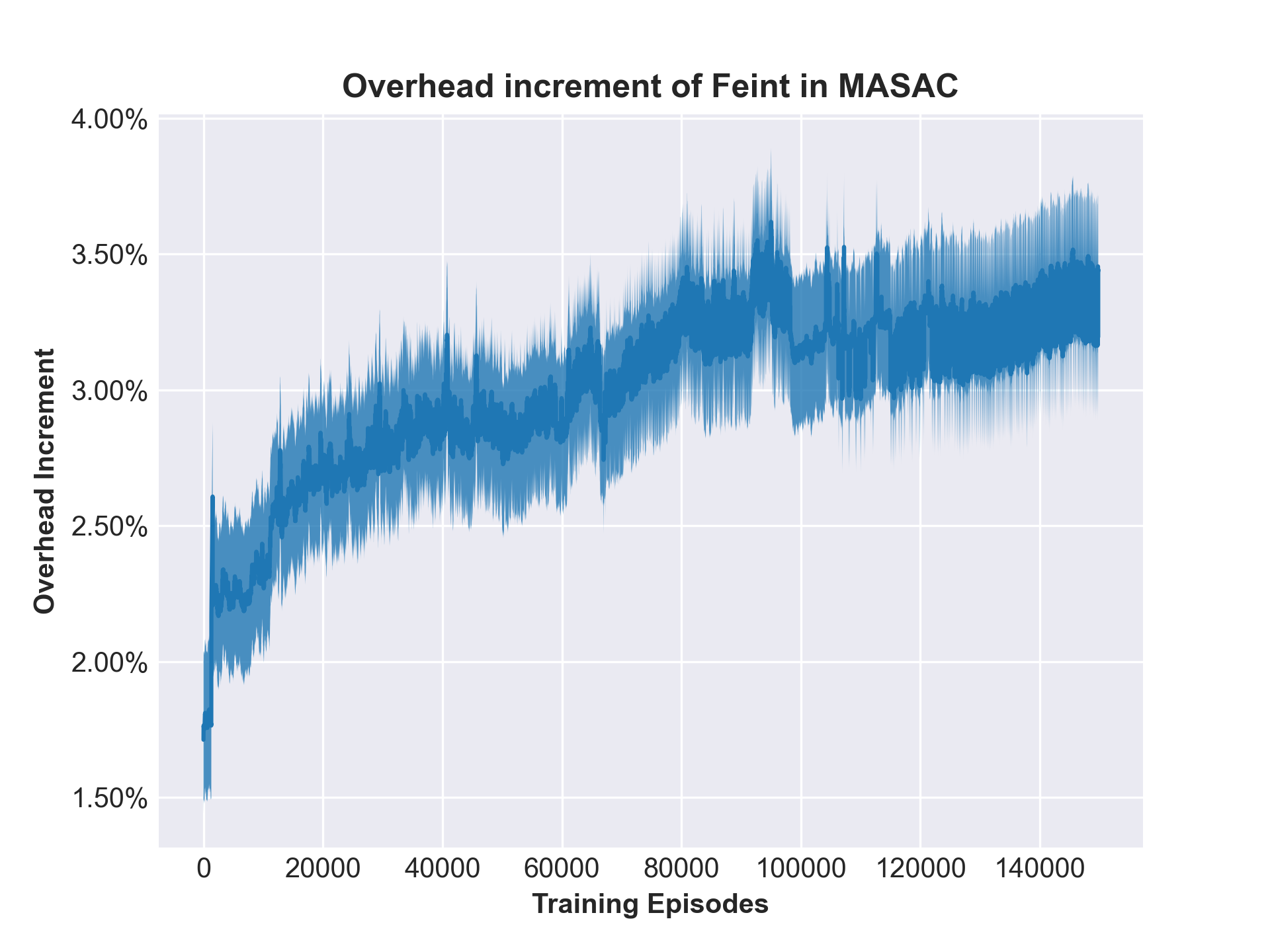}}\hfill
    \parbox{.24\linewidth}{\includegraphics[width=\linewidth]{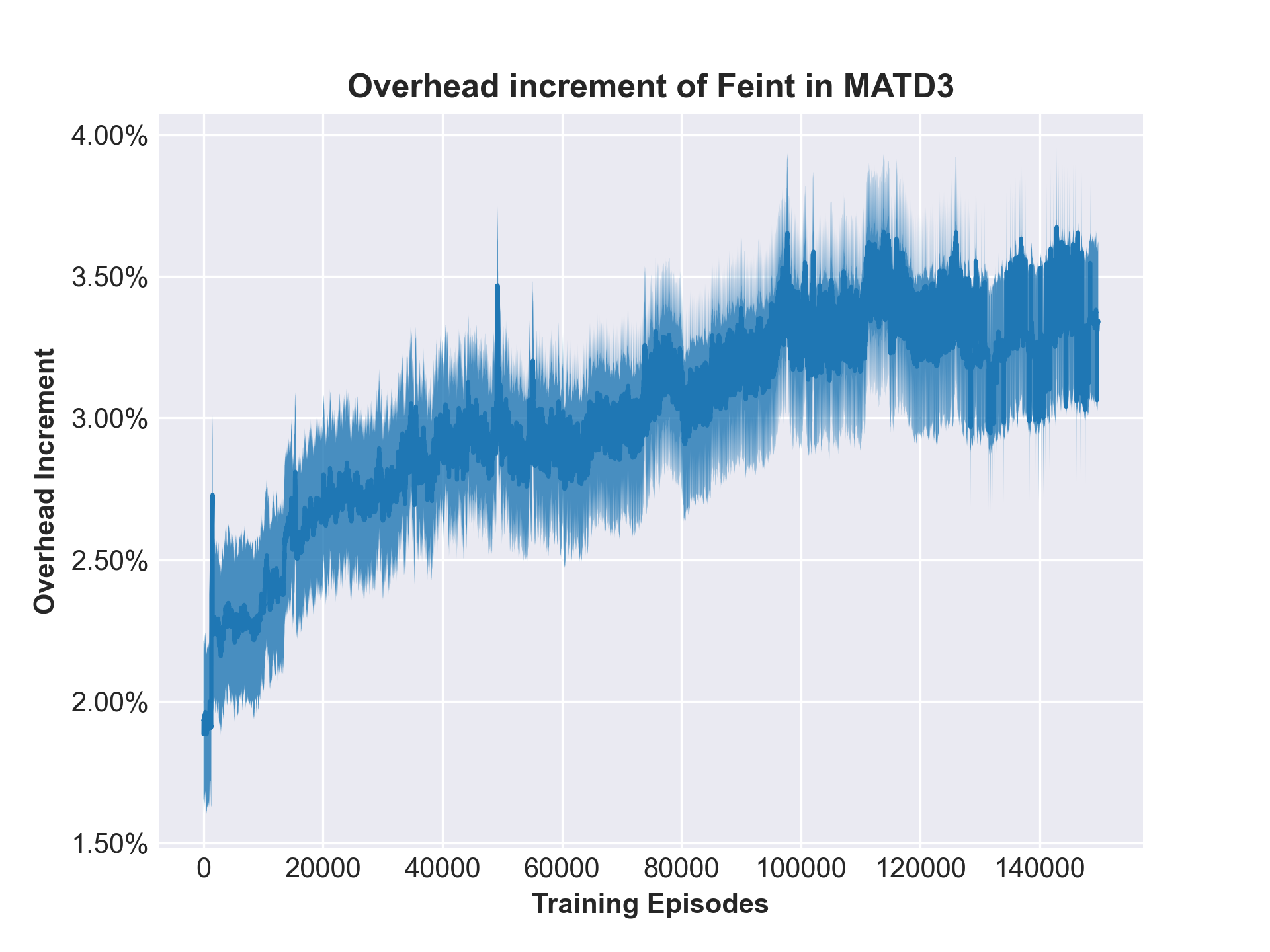}}\par
    \caption{Overhead of \feint the 1 VS 1 and 3 VS 3 scenarios using 4 MARL models.}
    \label{fig:experiment-overhead}
\end{figure}

Figure~\ref{fig:experiment-overhead} shows the results of our overhead analysis. We make two observations. First, fusing \feint in MARL training do incur some overhead increment in terms of running time. This is because the formalization and fusion of \feint in MARL incur additional calculation load. Secondly, in both MADDPG models and MAAC models, the increased overhead is generally lower than $5\%$, which still indicates that our proposed formalization of \feint actions can have enough feasibility and scalability on fusing with MARL models. Note that even we use two policy models for each agent in our implementation, our designs restrict that only one model is inferenced in each game step (Section~\ref{sec:implementation}), thus the overhead is low.

\section{Limitations}
\label{appendix:limitations}

We want to point out that our current design aims to provide a generalizable and as-automatic-as-possible
approach to concretely implement Feint behaviors into MARL models as a proof-of-concept. We acknowledge
that there are possible ways to optimize this template selection stage in different gaming scenarios, for example, adding human-knowledge guidance or learning heuristics. However, we believe that the variations at this
implementation level do not harm our main proof-of-concept contribution.

\end{document}